\documentclass[aps,prd,onecolumn,showpacs,showkeys,amsmath,amssymb]{revtex4}
\usepackage{graphicx}
\usepackage{dcolumn}
\usepackage{bm}
\begin{document}
%=====================================================================================
%=====================================================================================
\title{The $I^GJ^{PC}=1^-1^{-+}$ Tetraquark States}
%=====================================================================================
%=====================================================================================
%
\author{Hua-Xing Chen$^{1,2}$}
\email{hxchen@rcnp.osaka-u.ac.jp}
\author{Atsushi Hosaka$^{2}$}
\email{hosaka@rcnp.osaka-u.ac.jp}
\author{Shi-Lin Zhu$^{1}$}
\email{zhusl@phy.pku.edu.cn}
\affiliation{$^1$Department of Physics, Peking University, Beijing
100871, China \\ $^2$Research Center for Nuclear Physics, Osaka
University, Ibaraki 567--0047, Japan}
\begin{abstract}
We study the tetraquark states with $I^GJ^{PC}=1^-1^{-+}$ in the QCD
sum rule. After exhausting all possible flavor structures, we
analyses both the SVZ and finite energy sum rules. Both approaches
lead to a mass around 1.6 GeV for the state with the quark contents
$q q \bar q \bar q$, and around 2.0 GeV for the state with the quark
contents $q s \bar q \bar s$. The flavor structure $(\mathbf{\bar
3}\otimes \mathbf{\bar 6})\oplus(\mathbf{6}\otimes \mathbf{3})$ is
preferred. Our analysis strongly indicates that both $\pi_1(1600)$
and $\pi_1(2015)$ are also compatible with the exotic tetraquark
interpretation, which are sometimes labeled as candidates of the
$1^{-+}$ hybrid mesons. Moreover one of their dominant decay modes
is a pair of axial-vector and pseudoscalar mesons such as $b_1(1235)
\pi$, which is sometimes considered as the characteristic decay mode
of the hybrid mesons.
\end{abstract}
\maketitle
\pagenumbering{arabic}
%
%
%
%=====================================================================================
%=====================================================================================
\section{Introduction}\label{sec:intro}
%=====================================================================================
%=====================================================================================
%

Hadrons beyond the conventional quark model have been studied for
more than thirties years. For example, Jaffe suggested the low-lying
scalar mesons as good candidates of tetraquark states composed of
strongly correlated diquarks in 1976~\cite{Jaffe:1976ig}. Especially
there may exist some low-lying exotic mesons with quantum numbers
such as $(J^{PC})=(1^{-+})$ which $\bar q q$ mesons can not
access~\cite{Klempt:2007cp,Anikin:2005ur}. However the hybrid mesons
with explicit glue can carry such quantum numbers. The experimental
establishment of these states is a direct proof of the glue degree
of freedom in the low energy sector of QCD and of fundamental
importance.

The mass of the non-strange exotic hybrid meson from lattice QCD
simulations includes: 2GeV~\cite{McNeile:1998cp}, 1.74
GeV~\cite{Hedditch:2005zf}, and 1.8 GeV~\cite{Bernard:2003jd}. The
mass of its strange partner is 1.92 GeV~\cite{Hedditch:2005zf} and 2
GeV~\cite{Bernard:2003jd}. The hybrid meson mass from the
constituent glue model is 2 GeV~\cite{Iddir:2007dq} while the value
from the flux tube model is around 1.9
GeV~\cite{Isgur:1984bm,Page:1998gz}. The prediction from the QCD sum
rule approach is around 1.6 GeV~\cite{Jin:2002rw,Chetyrkin:2000tj}.
However, Yang obtained a surprisingly low mass around 1.26 GeV for
the $1^{-+}$ hybrid meson using QCD sum rule~\cite{Yang:2007cc}.

Up to now, there are several candidates of the exotic mesons with
$I^G(J^{PC})=1^-(1^{-+})$ experimentally. They are $\pi_1(1400)$,
$\pi_1(1600)$ and $\pi_1(2015)$. Their masses and widths are
($1376\pm17$, $300\pm40$) MeV, ($1653^{\Large +18}_{\Large -15}$,
$225^{\Large +45}_{\Large -28}$) MeV, ($2014 \pm 20 \pm 16$, $230
\pm 21 \pm 73$) MeV, respectively~\cite{Yao:2006px}. $\pi_1(1400)$
was observed in the reactions $\pi^- p \rightarrow \eta \pi^0
n$~\cite{Adams:2006sa}; $\bar p p \rightarrow \pi^0 \pi^0 \eta$ and
$\bar p n \rightarrow \pi^- \pi^0 \eta$~\cite{Abele:1999tf}; $\pi^-
p \rightarrow \eta \pi^- p$~\cite{Thompson:1997bs}. $\pi_1(1600)$
was observed in the reaction $\pi^- p \rightarrow \eta^\prime \pi^-
p$ ($\eta^\prime$ decays to $\eta \pi^+ \pi^-$ with a fraction
44.5\%)~\cite{Ivanov:2001rv}. Both $\pi_1(1600)$ and $\pi_1(2015)$
were observed in the reactions $\pi^- p \rightarrow \omega \pi^-
\pi^0 p$~\cite{Lu:2004yn} and $\pi^- p \rightarrow \eta \pi^+ \pi^-
\pi^- p$~\cite{Kuhn:2004en}. However, a more recent analysis of a
higher statistics sample from E852 $3\pi$ data found no evidence of
$\pi_1(1600)$ \cite{Dzierba:2005jg}. All the above observations were
from hadron-production experiments.

Recently, the CLAS Collaboration performed a photo-production
experiment to search for the $1^{-+}$ hybrid meson in the speculated
$3\pi$ final state in the charge exchange reaction $\gamma p
\rightarrow \pi^+ \pi^+ \pi^- (n)$~\cite{Nozar:2008be}. If
$\pi_1(1600)$ was an hybrid state, it was expected to be produced
with a strength near or much larger than 10\% of the $a_2(1320)$
meson from the theoretical models~\cite{theory}. However
$\pi_1(1600)$ was not observed with the expected strength. In fact
its production rate is less than 2\% of the $a_2(1320)$ meson. If
the $\pi_1(1600)$ signal from the hadron-production experiments is
not an artifact, the negative result of the photo-production
experiment suggests (1) either theoretical production rates are
overestimated significantly or (2) $\pi_1(1600)$ is a meson with a
different inner structure instead of a hybrid state.

In fact, the tetraquark states can also carry the exotic quantum
numbers $I^G(J^{PC})=1^-(1^{-+})$. It is important to note that the
gluon inside the hybrid meson can easily split into a pair of $q\bar
q$. Therefore tetraquarks can always have the same quantum numbers
as the hybrid mesons, including the exotic ones. Discovery of hadron
candidates with $J^{PC}=1^{-+}$ does not ensure that it is an exotic
hybrid meson. One has to exclude the other possibilities including
tetraquarks based on its mass, decay width and decay patterns etc.
This argument holds for all these claimed candidates of the hybrid
meson.

Tetraquark states in general have a richer internal structure than
ordinary $q \bar q $ states. For instance, a pair of quarks can be
in channels which can not be allowed in the ordinary hadrons. The
richness of the structure introduces complication in theoretical
studies. Therefore, one usually assumed one or a few particular
configurations which are motivated by some intuitions.

Recently, we have developed a systematic method for the study of
multiquark states in the QCD sum rule, and particular applications
have been made for several tetraquark
states~\cite{Chen:2006hy,Chen:2007xr,Chen:2008ej}. Our method is
essentially based on complete classification of independent
currents. By making suitable linear combinations of the independent
currents we can perform advanced analysis as compared with the
analysis of using only one type of current which limits the
potential of the OPE, and sometimes leads to unphysical results.

In this paper, we first classify the flavor structure of four-quark
system with quantum numbers $J^{PC} = 1^{-+}$. We find that there
are five iso-vector states. Then we construct tetraquark
interpolating currents by using both diquark-antidiquark
construction ($(q q)(\bar q \bar q)$) and quark-antiquark pairs ($(q
\bar q)(q \bar q)$). We verify that they are just different bases
and can be related to each other. Therefore they lead to the same
results. By using diquark-antidiquark currents, we perform the QCD
sum rule analysis, and calculate their masses. Our results suggest
that $\pi_1(1400)$ may not be explained by just using tetraquark
structure, and $\pi_1(1600)$ and $\pi_1(2015)$ could be explained by
the tetraquark mesons with quark contents $(qq)(\bar q \bar q)$ and
$(q s) (\bar q \bar s)$ respectively. The diquark and antidiquark
inside have a mixed flavor structure $(\mathbf{\bar
3}\otimes\mathbf{\bar 6}) \oplus (\mathbf{6}\otimes\mathbf{3})$.

This paper is organized as follows. In Sec.~\ref{sec:current}, we
construct the tetraquark currents using both diquark ($qq$) and
antidiquark ($\bar q \bar q$) currents. The tetraquark currents
constructed by using quark-antiquark ($\bar q q$) pairs are shown in
Appendix~\ref{app:mesoncurrent}. In Sec.~\ref{sec:svz}, we perform a
QCD sum rule analysis by using these currents, and calculate their
OPEs. In Sec.~\ref{sec:numeri}, the numerical result is obtained for
their masses. In Sec.~\ref{sec:fesr}, we use finite energy sum rule
to calculate their masses again. We discuss the decay patterns of
these $1^{-+}$ tetraquark states in Sec.~\ref{sec:decay}.
Sec.~\ref{sec:summary} is a summary.

%
%=====================================================================================
%=====================================================================================
\section{Tetraquark Currents}\label{sec:current}
%=====================================================================================
%=====================================================================================
%

\begin{figure}[hbt]
\begin{center}
\scalebox{1}{\includegraphics{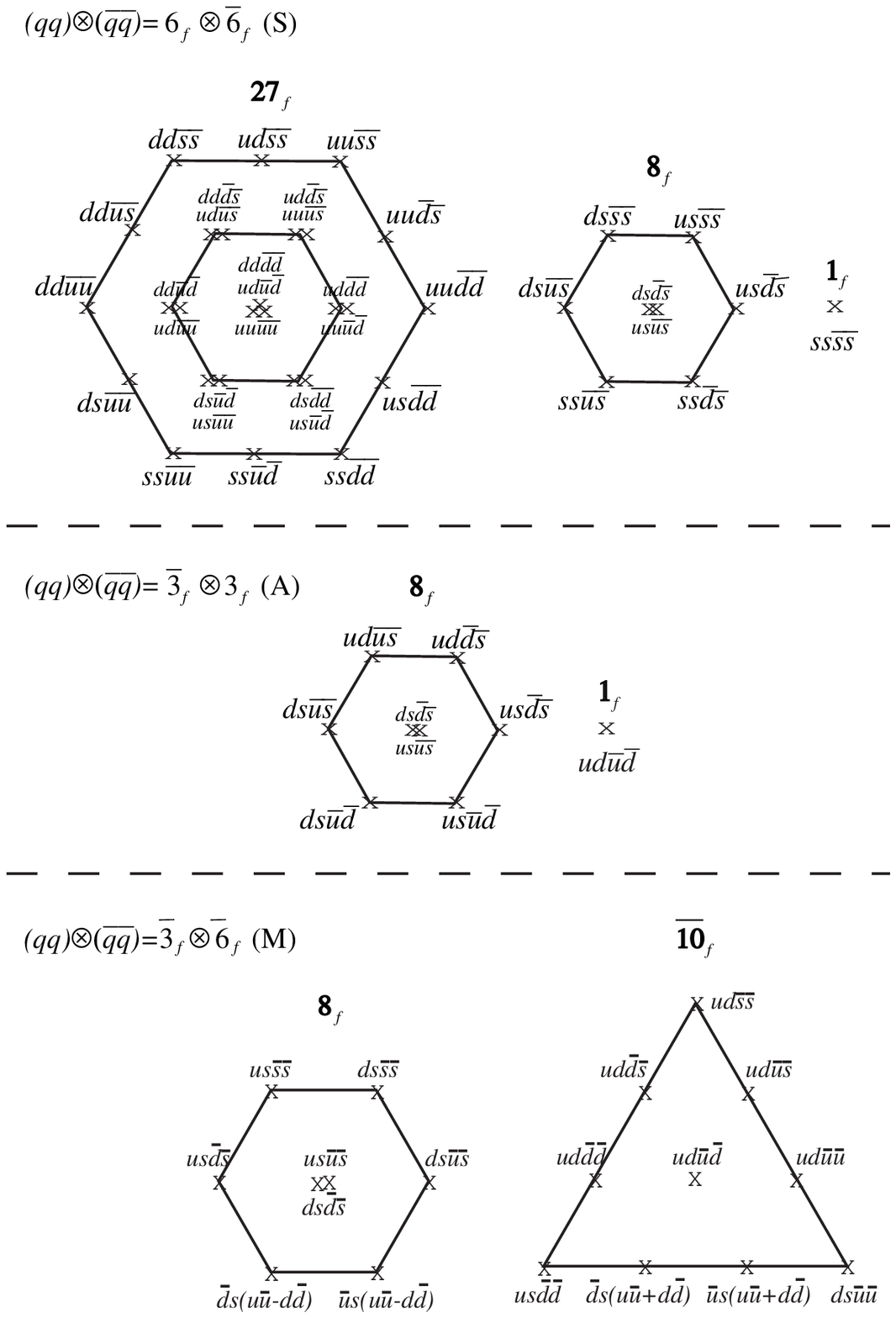}} \caption{Weight diagrams
for $\mathbf{6_f}\otimes\mathbf{\bar 6_f} (\mathbf{S})$ (top panel),
$\mathbf{\bar 3_f}\otimes\mathbf{3_f} (\mathbf{A})$ (middle panel),
and $\mathbf{\bar 3_f} \otimes \mathbf{\bar 6_f} (\mathbf{M})$
(bottom panel). The weight diagram for
$\mathbf{6_f}\otimes\mathbf{3_f} (\mathbf{M})$ is the
charge-conjugation transformation of the bottom one.}
\label{fig:tetra}
\end{center}
\end{figure}

In order to construct proper tetraquark currents, let us start with
the consideration of the charge-conjugation symmetry. The
charge-conjugation transformation changes diquarks into
antidiquarks, while it maintains their flavor structures. If a
tetraquark state has a definite charge-conjugation parity, either
positive or negative, the internal diquark ($qq$) and antidiquark
($\bar q \bar q$) must have the same flavor symmetry, which is
either symmetric flavor structure $\mathbf{6_f} \otimes \mathbf{\bar
6_f}$ ($\mathbf{S}$) or antisymmetric flavor structure $\mathbf{\bar
3_f} \otimes \mathbf{3_f}$ ($\mathbf{A}$), and can not have mixed
flavor symmetry neither $\mathbf{\bar 3_f} \otimes \mathbf{\bar
6_f}$ nor $\mathbf{6_f} \otimes \mathbf{3_f}$ ($\mathbf{M}$).
However, combinations of $\mathbf{\bar 3_f} \otimes \mathbf{\bar
6_f}$ and $\mathbf{6_f} \otimes \mathbf{3_f}$ can have a definite
charge-conjugation parity. Therefore, in order to study the
tetraquark state of $I^GJ^{PC} =1^- 1^{-+}$, we need to consider the
following structures of currents
\begin{eqnarray} \nonumber
q q\bar q \bar q (\mathbf{S})\, , q s\bar q \bar s (\mathbf{S})
&\sim& \mathbf{6_f} \otimes \mathbf{\bar 6_f}~~~(\mathbf{S}) \, ,
\\ \nonumber q s\bar q \bar s (\mathbf{A}) &\sim& \mathbf{\bar 3_f} \otimes \mathbf{3_f}~~~(\mathbf{A}) \, ,
\\ \nonumber q q \bar q \bar q (\mathbf{M}) \, , q s \bar q \bar s (\mathbf{M})
&\sim& (\mathbf{\bar 3_f} \otimes \mathbf{\bar 6_f}) \oplus
(\mathbf{6_f} \otimes \mathbf{3_f})~~~(\mathbf{M}) \, ,
\end{eqnarray}
where $q$ represents an $up$ or $down$ quark, and $s$ represents a
$strange$ quark. The flavor structures are shown in
Fig.~\ref{fig:tetra} in terms of $SU(3)$ weight diagrams. The quark
contents indicated at vertices follow the ideal mixing scheme for
inner vertices where the mixing is allowed. In the $SU(3)$ limit,
the quark contents are suitable combinations of the ones shown in
this figures. However, the $strange$ quark has a significantly
larger mass than $up$ and $down$ quarks (current quark mass), and
so, the ideal mixing is expected to work well for hadrons except for
pseudoscalar mesons. The flavor structure in the ideal mixing is
also simpler than that in the $SU(3)$ limit. Therefore, we will use
the ideal mixing in our QCD sum rule studies.
%
%%%%%%%%%%%%%%%%%%%%%%%%%%%%%%%%%%%%%%%%%%%%%%%%%%%%%%%%%%%%%%%%%%%%%%%%%%%%%%
%---------figure 6*6
%%%%%%%%%%%%%%%%%%%%%%%%%%%%%%%%%%%%%%%%%%%%%%%%%%%%%%%%%%%%%%%%%%%%%%%%%%%%%%
%

In the following subsections, we first construct currents by using
diquark ($qq$) and antidiquark ($\bar q \bar q$) currents, and then
we show the currents with explicit quark contents. The currents
constructed by using quark-antiquark ($\bar q q$) pairs can be
related to these diquark currents, and are shown in the
Appendix.~\ref{app:mesoncurrent}. The tensor currents
$\eta_{\mu\nu}$ ($\eta_{\mu\nu} = - \eta_{\nu\mu}$) can also have
$I^GJ^{PC} = 1^-1^{-+}$. By using tensor currents, we obtain the
similar results, which will be shown in our future work.

%
%=====================================================================================
%=====================================================================================
\subsection{$(qq)(\bar q \bar q)$
Currents}\label{subsec:diquarkcurrent}
%=====================================================================================
%=====================================================================================
%

We attempt to construct the tetraquark currents using diquark ($qq$)
and antidiquark ($\bar q \bar q$) currents. For each state having
the symmetric flavor structure $\mathbf{6_f} \otimes \mathbf{\bar
6_f}$ ($\mathbf{S}$), there are two $(qq)(\bar q \bar q)$ currents
of $J^{PC}=1^{-+}$, which are independent
%
%%%%%%%%%%%%%%%%%%%%%%%%%%%%%%%%%%%%%%%%%%%%%%%%%%%%%%%%%%%%%%%%%%%%%%%%%%%%%%
\begin{eqnarray}\label{def:currentS}
%-------------------------------------psi S 1------------------------------------
\psi^S_{1\mu} &=& q_{1a}^T C \gamma_5 q_{2b} (\bar{q}_{3a}
\gamma_\mu \gamma_5 C \bar{q}_{4b}^T + \bar{q}_{3b} \gamma_\mu
\gamma_5 C \bar{q}_{4a}^T) + q_{1a}^T C \gamma_\mu \gamma_5 q_{2b}
(\bar{q}_{3a} \gamma_5 C \bar{q}_{4b}^T + \bar{q}_{3b} \gamma_5 C
\bar{q}_{4a}^T) \, ,
%-------------------------------------psi S 2------------------------------------
\\ \nonumber \psi^S_{2\mu} &=& q_{1a}^T C \gamma^\nu
q_{2b} (\bar{q}_{3a} \sigma_{\mu\nu} C \bar{q}_{4b}^T - \bar{q}_{3b}
\sigma_{\mu\nu} C \bar{q}_{4a}^T) + q_{1a}^T C \sigma_{\mu\nu}
q_{2b} (\bar{q}_{3a} \gamma^\nu C \bar{q}_{4b}^T - \bar{q}_{3b}
\gamma^\nu C \bar{q}_{4a}^T) \, ,
\end{eqnarray}
%%%%%%%%%%%%%%%%%%%%%%%%%%%%%%%%%%%%%%%%%%%%%%%%%%%%%%%%%%%%%%%%%%%%%%%%%%%%%%
%
where the sum over repeated indices ($\mu$, $\nu, \cdots$ for Dirac
spinor indices, and $a, b, \cdots$ for color indices) is taken. $C$
is the charge-conjugation matrix, $q_1$ and $q_2$ represent quarks,
and $q_3$ and $q_4$ represent antiquarks. For the antisymmetry
flavor structure $\mathbf{\bar 3_f} \otimes \mathbf{3_f}$
($\mathbf{A}$), we also find that there are two independent $(q
q)(\bar q \bar q)$ currents,
%
%%%%%%%%%%%%%%%%%%%%%%%%%%%%%%%%%%%%%%%%%%%%%%%%%%%%%%%%%%%%%%%%%%%%%%%%%%%%%%
\begin{eqnarray}\label{def:currentA}
%-------------------------------------psi A 1------------------------------------
\psi^A_{1\mu} &=& q_{1a}^T C \gamma_5 q_{2b} (\bar{q}_{3a}
\gamma_\mu \gamma_5 C \bar{q}_{4b}^T -\bar{q}_{3b} \gamma_\mu
\gamma_5 C \bar{q}_{4a}^T) + q_{1a}^T C \gamma_\mu \gamma_5 q_{2b}
(\bar{q}_{3a} \gamma_5 C \bar{q}_{4b}^T - \bar{q}_{3b} \gamma_5 C
\bar{q}_{4a}^T) \, ,
%-------------------------------------psi A 2------------------------------------
\\ \nonumber \psi^A_{2\mu} &=&  q_{1a}^T C \gamma^\nu q_{2b} (\bar{q}_{3a}
\sigma_{\mu\nu} C \bar{q}_{4b}^T + \bar{q}_{3b} \sigma_{\mu\nu} C
\bar{q}_{4a}^T) + q_{1a}^T C \sigma_{\mu\nu} q_{2b} (\bar{q}_{3a}
\gamma^\nu C \bar{q}_{4b}^T + \bar{q}_{3b} \gamma^\nu C
\bar{q}_{4a}^T) \, ,
\end{eqnarray}
%%%%%%%%%%%%%%%%%%%%%%%%%%%%%%%%%%%%%%%%%%%%%%%%%%%%%%%%%%%%%%%%%%%%%%%%%%%%%%
%

For each state containing diquark and antidiquark having either the
flavor structure $\mathbf{\bar 3_f} \otimes \mathbf{\bar 6_f}$ or
$\mathbf{6_f} \otimes \mathbf{3_f}$, there are no currents of
quantum numbers $J^{PC}=1^{-+}$. However, their combinations
$(\mathbf{\bar 3_f} \otimes \mathbf{\bar 6_f} )\oplus (\mathbf{6_f}
\otimes \mathbf{3_f})$ can have the quantum numbers $J^{PC}=1^{-+}$.
We first define the currents $\psi^{ML}_{i\mu}$ which belong to the
flavor representation $\mathbf{\bar 3_f} \otimes \mathbf{\bar 6_f}$,
and the currents $\psi^{MR}_{i\mu}$ which belong to the flavor
representation $\mathbf{6_f} \otimes \mathbf{3_f}$ separately. We
find the following four independent currents:
%
%%%%%%%%%%%%%%%%%%%%%%%%%%%%%%%%%%%%%%%%%%%%%%%%%%%%%%%%%%%%%%%%%%%%%%%%%%%%%%
\begin{eqnarray}
%-------------------------------------psi ML 1------------------------------------
\nonumber \psi^{ML}_{1\mu} &=& q_{1a}^T C \gamma_\mu q_{2b}
(\bar{q}_{3a} C \bar{q}_{4b}^T + \bar{q}_{3b} C \bar{q}_{4a}^T) \, ,
%-------------------------------------psi ML 2------------------------------------
\\ \nonumber \psi^{ML}_{2\mu} &=& q_{1a}^T C \sigma_{\mu\nu} \gamma_5
q_{2b} (\bar{q}_{3a} \gamma^{\nu} \gamma_5 C \bar{q}_{4b}^T +
\bar{q}_{3b} \gamma^{\nu} \gamma_5 C \bar{q}_{4a}^T) \, ,
%-------------------------------------psi ML 3------------------------------------
\\ \nonumber \psi^{ML}_{3\mu} &=& q_{1a}^T C q_{2b} (\bar{q}_{3a} \gamma_\mu C
\bar{q}_{4b}^T - \bar{q}_{3b} \gamma_\mu C \bar{q}_{4a}^T) \, ,
%-------------------------------------psi ML 4------------------------------------
\\ \nonumber \psi^{ML}_{4\mu} &=& q_{1a}^T C \gamma^{\nu} \gamma_5
q_{2b} (\bar{q}_{3a} \sigma_{\mu\nu} \gamma_5 C \bar{q}_{4b}^T -
\bar{q}_{3b} \sigma_{\mu\nu} \gamma_5 C \bar{q}_{4a}^T) \, ,
\end{eqnarray}
%%%%%%%%%%%%%%%%%%%%%%%%%%%%%%%%%%%%%%%%%%%%%%%%%%%%%%%%%%%%%%%%%%%%%%%%%%%%%%
%
%
%%%%%%%%%%%%%%%%%%%%%%%%%%%%%%%%%%%%%%%%%%%%%%%%%%%%%%%%%%%%%%%%%%%%%%%%%%%%%%
\begin{eqnarray}
%-------------------------------------psi MR 1------------------------------------
\nonumber \psi^{MR}_{1\mu} &=& q_{1a}^T C q_{2b} (\bar{q}_{3a}
\gamma_\mu C \bar{q}_{4b}^T + \bar{q}_{3b} \gamma_\mu C
\bar{q}_{4a}^T) \, ,
%-------------------------------------psi ML 2------------------------------------
\\ \nonumber \psi^{MR}_{2\mu} &=& q_{1a}^T C \gamma^{\nu} \gamma_5
q_{2b} (\bar{q}_{3a} \sigma_{\mu\nu} \gamma_5 C \bar{q}_{4b}^T +
\bar{q}_{3b} \sigma_{\mu\nu} \gamma_5 C \bar{q}_{4a}^T) \, ,
%-------------------------------------psi ML 3------------------------------------
\\ \nonumber \psi^{MR}_{3\mu} &=& q_{1a}^T C \gamma_\mu q_{2b} (\bar{q}_{3a} C
\bar{q}_{4b}^T - \bar{q}_{3b} C \bar{q}_{4a}^T) \, ,
%-------------------------------------psi ML 4------------------------------------
\\ \nonumber \psi^{MR}_{4\mu} &=& q_{1a}^T C \sigma_{\mu\nu} \gamma_5
q_{2b} (\bar{q}_{3a} \gamma^{\nu} \gamma_5 C \bar{q}_{4b}^T -
\bar{q}_{3b} \gamma^{\nu} \gamma_5 C \bar{q}_{4a}^T) \, .
\end{eqnarray}
%%%%%%%%%%%%%%%%%%%%%%%%%%%%%%%%%%%%%%%%%%%%%%%%%%%%%%%%%%%%%%%%%%%%%%%%%%%%%%
%
They all have quantum numbers $J^P=1^-$ but no good
charge-conjugation parity. However, their mixing can have a definite
charge-conjugation parity,
%
%%%%%%%%%%%%%%%%%%%%%%%%%%%%%%%%%%%%%%%%%%%%%%%%%%%%%%%%%%%%%%%%%%%%%%%%%%%%%%
\begin{eqnarray}\label{def:currentM}
%-------------------------------------psi M------------------------------------
\psi^M_{i\mu} &=& \psi^{ML}_{i\mu} \pm \psi^{MR}_{i\mu} \, ,
\end{eqnarray}
%%%%%%%%%%%%%%%%%%%%%%%%%%%%%%%%%%%%%%%%%%%%%%%%%%%%%%%%%%%%%%%%%%%%%%%%%%%%%%
%
where the $+$ and $-$ combinations correspond to the
charge-conjugation parity positive and negative, respectively. In
the present work, we only consider the positive one.

%
%=====================================================================================
%=====================================================================================
\subsection{Iso-Vector Currents}\label{subsec:isoveccurrent}
%=====================================================================================
%=====================================================================================
%

For the study of the present exotic tetraquark state, we need to
construct iso-vector ($I=1$) currents. There are two isospin
triplets belonging to the flavor representation $\mathbf{6}_f
\otimes \mathbf{\bar 6}_f$, one isospin triplet belonging to the
flavor representation $\mathbf{\bar 3}_f \otimes \mathbf{3}_f$, and
two isospin triplets belonging to the flavor representation
$(\mathbf{\bar 3}_f \otimes \mathbf{\bar 6}_f)\oplus(\mathbf{6}_f
\otimes \mathbf{3}_f)$ (Fig.~\ref{fig:tetra}). For each state, there
are several independent currents. We list them in the following.
\begin{enumerate}

\item For the two isospin
triplets belonging to $\mathbf{6}_f \otimes \mathbf{\bar 6}_f$
($\mathbf{S}$):
%
%%%%%%%%%%%%%%%%%%%%%%%%%%%%%%%%%%%%%%%%%%%%%%%%%%%%%%%%%%%%%%%%%%%%%%%%%%%%%%
\begin{eqnarray}\label{def:Scurrent}
\nonumber && \left \{
\begin{array}{l}
\eta^S_{1\mu} \equiv \psi_{1\mu}^S(qq\bar q \bar q) \sim u_a^T C
\gamma_5 d_b (\bar{u}_a \gamma_\mu \gamma_5 C \bar{d}_b^T +
\bar{u}_b \gamma_\mu \gamma_5 C \bar{d}_a^T) + u_a^T C \gamma_\mu
\gamma_5 d_b (\bar{u}_a \gamma_5 C \bar{d}_b^T + \bar{u}_b \gamma_5
C \bar{d}_a^T) \, ,
\\ \eta^S_{2\mu} \equiv \psi_{2\mu}^S(qq\bar q \bar q)
\sim u_a^T C \gamma^\nu d_b (\bar{u}_a \sigma_{\mu\nu} C \bar{d}_b^T
- \bar{u}_b \sigma_{\mu\nu} C \bar{d}_a^T) + u_a^T C \sigma_{\mu\nu}
d_b (\bar{u}_a \gamma^\nu C \bar{d}_b^T - \bar{u}_b \gamma^\nu C
\bar{d}_a^T) \, ,
\end{array} \right.
\\ \nonumber && \left \{
\begin{array}{l} \eta^S_{3\mu} \equiv \psi_{1\mu}^S(q s \bar q \bar s) \sim u_a^T C \gamma_5 s_b
(\bar{u}_a \gamma_\mu \gamma_5 C \bar{s}_b^T + \bar{u}_b \gamma_\mu
\gamma_5 C \bar{s}_a^T) + u_a^T C \gamma_\mu \gamma_5 s_b (\bar{u}_a
\gamma_5 C \bar{s}_b^T + \bar{u}_b \gamma_5 C \bar{s}_a^T) \, ,
\\ \eta^S_{4\mu} \equiv \psi_{2\mu}^S(q s \bar q \bar s)
\sim u_a^T C \gamma^\nu s_b (\bar{u}_a \sigma_{\mu\nu} C \bar{s}_b^T
- \bar{u}_b \sigma_{\mu\nu} C \bar{s}_a^T) + u_a^T C \sigma_{\mu\nu}
s_b (\bar{u}_a \gamma^\nu C \bar{s}_b^T - \bar{u}_b \gamma^\nu C
\bar{s}_a^T) \, .
\end{array} \right.
\end{eqnarray}
%%%%%%%%%%%%%%%%%%%%%%%%%%%%%%%%%%%%%%%%%%%%%%%%%%%%%%%%%%%%%%%%%%%%%%%%%%%%%%
%
where $\eta^S_{1\mu}$ and $\eta^S_{2\mu}$ are the two independent
currents containing only light flavors, and $\eta^S_{3\mu}$ and
$\eta^S_{4\mu}$ are the two independent ones containing one $s \bar
s$ quark pair.

\item For the isospin
triplet belonging to $\mathbf{\bar 3}_f \otimes \mathbf{3}_f$
($\mathbf{A}$):
%
%%%%%%%%%%%%%%%%%%%%%%%%%%%%%%%%%%%%%%%%%%%%%%%%%%%%%%%%%%%%%%%%%%%%%%%%%%%%%%
\begin{eqnarray}\label{def:Mcurrent}
&& \nonumber \left \{
\begin{array}{l}
\eta^A_{1\mu} \equiv \psi_{1\mu}^A(q s\bar q \bar s) \sim u_a^T C
\gamma_5 s_b (\bar{u}_a \gamma_\mu \gamma_5 C \bar{s}_b^T -\bar{u}_b
\gamma_\mu \gamma_5 C \bar{s}_a^T) + u_a^T C \gamma_\mu \gamma_5 s_b
(\bar{u}_a \gamma_5 C \bar{s}_b^T - \bar{u}_b \gamma_5 C
\bar{s}_a^T) \, ,
\\ \eta^A_{2\mu} \equiv \psi_{2\mu}^A(q s\bar q \bar s)
\sim u_a^T C \gamma^\nu s_b (\bar{u}_a \sigma_{\mu\nu} C \bar{s}_b^T
+ \bar{u}_b \sigma_{\mu\nu} C \bar{s}_a^T) + u_a^T C \sigma_{\mu\nu}
s_b (\bar{u}_a \gamma^\nu C \bar{s}_b^T + \bar{u}_b \gamma^\nu C
\bar{s}_a^T) \, ,
\end{array} \right.
\end{eqnarray}
where $\eta^A_{1\mu}$ and $\eta^A_{2\mu}$ are the two independent
currents.

\item For the two isospin
triplets belonging to $(\mathbf{\bar 3}_f \otimes \mathbf{\bar
6}_f)\oplus(\mathbf{6}_f \otimes \mathbf{3}_f)$ ($\mathbf{M}$):
\begin{eqnarray}\label{def:Mcurrent}
&& \nonumber \left \{
\begin{array}{l}
\eta^M_{1\mu} \equiv \psi^{M}_{1\mu}(q q \bar q \bar q) \sim u_{a}^T
C \gamma_\mu d_{b} (\bar{u}_{a} C \bar{d}_{b}^T + \bar{u}_{b} C
\bar{d}_{a}^T) + u_{a}^T C d_{b} (\bar{u}_{a} \gamma_\mu C
\bar{d}_{b}^T + \bar{u}_{b} \gamma_\mu C \bar{d}_{a}^T) \, ,
\\ \eta^M_{2\mu} \equiv \psi^{M}_{2\mu}(q q \bar q \bar q) \sim u_{a}^T C \sigma_{\mu\nu} \gamma_5
d_{b} (\bar{u}_{a} \gamma^{\nu} \gamma_5 C \bar{d}_{b}^T +
\bar{u}_{b} \gamma^{\nu} \gamma_5 C \bar{d}_{a}^T) + u_{a}^T C
\gamma^{\nu} \gamma_5 d_{b} (\bar{u}_{a} \sigma_{\mu\nu} \gamma_5 C
\bar{d}_{b}^T + \bar{u}_{b} \sigma_{\mu\nu} \gamma_5 C
\bar{d}_{a}^T) \, ,
\\ \eta^M_{3\mu} \equiv \psi^{M}_{3\mu}(q q \bar q \bar q) \sim u_{a}^T C d_{b} (\bar{u}_{a} \gamma_\mu C
\bar{d}_{b}^T - \bar{u}_{b} \gamma_\mu C \bar{d}_{a}^T) + u_{a}^T C
\gamma_\mu d_{b} (\bar{u}_{a} C \bar{d}_{b}^T - \bar{u}_{b} C
\bar{d}_{a}^T) \, ,
\\ \eta^M_{4\mu} \equiv \psi^{M}_{4\mu}(q q \bar q \bar q) \sim u_{a}^T C \gamma^{\nu} \gamma_5
d_{b} (\bar{u}_{a} \sigma_{\mu\nu} \gamma_5 C \bar{d}_{b}^T -
\bar{u}_{b} \sigma_{\mu\nu} \gamma_5 C \bar{d}_{a}^T) + u_{a}^T C
\sigma_{\mu\nu} \gamma_5 d_{b} (\bar{u}_{a} \gamma^{\nu} \gamma_5 C
\bar{d}_{b}^T - \bar{u}_{b} \gamma^{\nu} \gamma_5 C \bar{d}_{a}^T)
\, ,
\end{array} \right.
\\ \nonumber && \left \{
\begin{array}{l}
\eta^M_{5\mu} \equiv \psi^{M}_{1\mu}(q s \bar q \bar s) \sim u_{a}^T
C \gamma_\mu s_{b} (\bar{u}_{a} C \bar{s}_{b}^T + \bar{u}_{b} C
\bar{s}_{a}^T) + u_{a}^T C s_{b} (\bar{u}_{a} \gamma_\mu C
\bar{s}_{b}^T + \bar{u}_{b} \gamma_\mu C \bar{s}_{a}^T) \, ,
\\ \eta^M_{6\mu} \equiv \psi^{M}_{2\mu}(q s \bar q \bar s) \sim u_{a}^T C \sigma_{\mu\nu} \gamma_5
s_{b} (\bar{u}_{a} \gamma^{\nu} \gamma_5 C \bar{s}_{b}^T +
\bar{u}_{b} \gamma^{\nu} \gamma_5 C \bar{s}_{a}^T) + u_{a}^T C
\gamma^{\nu} \gamma_5 s_{b} (\bar{u}_{a} \sigma_{\mu\nu} \gamma_5 C
\bar{s}_{b}^T + \bar{u}_{b} \sigma_{\mu\nu} \gamma_5 C
\bar{s}_{a}^T) \, ,
\\ \eta^M_{7\mu} \equiv \psi^{M}_{3\mu}(q s \bar q \bar s) \sim u_{a}^T C s_{b} (\bar{u}_{a} \gamma_\mu C
\bar{s}_{b}^T - \bar{u}_{b} \gamma_\mu C \bar{s}_{a}^T) + u_{a}^T C
\gamma_\mu s_{b} (\bar{u}_{a} C \bar{s}_{b}^T - \bar{u}_{b} C
\bar{s}_{a}^T) \, ,
\\ \eta^M_{8\mu} \equiv \psi^{M}_{4\mu}(q s \bar q \bar s) \sim u_{a}^T C \gamma^{\nu} \gamma_5
s_{b} (\bar{u}_{a} \sigma_{\mu\nu} \gamma_5 C \bar{s}_{b}^T -
\bar{u}_{b} \sigma_{\mu\nu} \gamma_5 C \bar{s}_{a}^T) + u_{a}^T C
\sigma_{\mu\nu} \gamma_5 s_{b} (\bar{u}_{a} \gamma^{\nu} \gamma_5 C
\bar{s}_{b}^T - \bar{u}_{b} \gamma^{\nu} \gamma_5 C \bar{s}_{a}^T)
\, ,
\end{array} \right.
\end{eqnarray}
%%%%%%%%%%%%%%%%%%%%%%%%%%%%%%%%%%%%%%%%%%%%%%%%%%%%%%%%%%%%%%%%%%%%%%%%%%%%%%
%
where $\eta^M_{1,2,3,4}$ are the four independent currents
containing only light flavors, and $\eta^M_{1,2,3,4}$ are the four
independent ones containing one $s \bar s$ quark pair.

\end{enumerate}
We use $\sim$ to make clear that the quark contents here are not
exactly correct. For instance, in the current $\eta^A_{1\mu}$, the
state $us\bar u \bar s$ does not have isospin one. The correct quark
contents should be $(u s \bar u \bar s - d s \bar d \bar s)$.
However, in the following QCD sum rule analysis, we shall not
include the mass of $up$ and $down$ quarks and choose the same value
for $\langle \bar u u \rangle$ and $\langle \bar d d \rangle$.
Therefore, the QCD sum rule results for $\eta^A_1$ with quark
contents $u s \bar u \bar s$ and $(u s \bar u \bar s - d s \bar d
\bar s)$ are the same.

%
%=====================================================================================
%=====================================================================================
\section{SVZ sum rule}\label{sec:svz}
%=====================================================================================
%=====================================================================================
%

For the past decades QCD sum rule has proven to be a very powerful
and successful non-perturbative
method~\cite{Shifman:1978bx,Reinders:1984sr}. In sum rule analyses,
we consider two-point correlation functions:
%
%%%%%%%%%%%%%%%%%%%%%%%%%%%%%%%%%%%%%%%%%%%%%%%%%%%%%%%%%%%%%%%%%%%%%%%%%%%%%%
\begin{equation}
\Pi_{\mu\nu}(q^2) \, \equiv \, i \int d^4x e^{iqx} \langle 0 | T
\eta_\mu(x) { \eta_\nu^\dagger } (0) | 0 \rangle \, , \label{def:pi}
\end{equation}
%%%%%%%%%%%%%%%%%%%%%%%%%%%%%%%%%%%%%%%%%%%%%%%%%%%%%%%%%%%%%%%%%%%%%%%%%%%%%%
%
where $\eta_\mu$ is an interpolating current for the tetraquark. The
Lorentz structure can be simplified to be:
%
%%%%%%%%%%%%%%%%%%%%%%%%%%%%%%%%%%%%%%%%%%%%%%%%%%%%%%%%%%%%%%%%%%%%%%%%%%%%%%
\begin{equation}
\Pi_{\mu\nu}(q^2) = ( {q_\mu q_\nu \over q^2} - g_{\mu\nu} )
\Pi^{(1)}(q^2) + {q_\mu q_\nu \over q^2} \Pi^{(0)}(q^2) \, .
\label{def:pi1}
\end{equation}
%%%%%%%%%%%%%%%%%%%%%%%%%%%%%%%%%%%%%%%%%%%%%%%%%%%%%%%%%%%%%%%%%%%%%%%%%%%%%%
%

We compute $\Pi(q^2)$ in the operator product expansion (OPE) of QCD
up to certain order in the expansion, which is then matched with a
hadronic parametrization to extract information of hadron
properties. At the hadron level, we express the correlation function
in the form of the dispersion relation with a spectral function:
%
%%%%%%%%%%%%%%%%%%%%%%%%%%%%%%%%%%%%%%%%%%%%%%%%%%%%%%%%%%%%%%%%%%%%%%%%%%%%%%
\begin{equation}
\Pi^{(1)}(q^2)=\int^\infty_{s_<}\frac{\rho(s)}{s-q^2-i\varepsilon}ds
\, , \label{eq:disper}
\end{equation}
%%%%%%%%%%%%%%%%%%%%%%%%%%%%%%%%%%%%%%%%%%%%%%%%%%%%%%%%%%%%%%%%%%%%%%%%%%%%%%
%
where the integration starts from the mass square of all current
quarks. The the spectral density $\rho(s)$ is defined to be
%
%%%%%%%%%%%%%%%%%%%%%%%%%%%%%%%%%%%%%%%%%%%%%%%%%%%%%%%%%%%%%%%%%%%%%%%%%%%%%%
\begin{eqnarray}
\rho(s) & \equiv & \sum_n\delta(s-M^2_n)\langle
0|\eta|n\rangle\langle n|{\eta^\dagger}|0\rangle \ \nonumber\\ &=&
f^2_Y\delta(s-M^2_Y)+ \rm{higher\,\,states}\, . \label{eq:rho}
\end{eqnarray}
%%%%%%%%%%%%%%%%%%%%%%%%%%%%%%%%%%%%%%%%%%%%%%%%%%%%%%%%%%%%%%%%%%%%%%%%%%%%%%
%
For the second equation, as usual, we adopt a parametrization of one
pole dominance for the ground state $Y$ and a continuum
contribution. The sum rule analysis is then performed after the
Borel transformation of the two expressions of the correlation
function, (\ref{def:pi}) and (\ref{eq:disper})
%
%%%%%%%%%%%%%%%%%%%%%%%%%%%%%%%%%%%%%%%%%%%%%%%%%%%%%%%%%%%%%%%%%%%%%%%%%%%%%%
\begin{equation}
\Pi^{(all)}(M_B^2)\equiv\mathcal{B}_{M_B^2}\Pi^{(1)}(p^2)=\int^\infty_{s_<}
e^{-s/M_B^2} \rho(s)ds \, . \label{eq:borel}
\end{equation}
%%%%%%%%%%%%%%%%%%%%%%%%%%%%%%%%%%%%%%%%%%%%%%%%%%%%%%%%%%%%%%%%%%%%%%%%%%%%%%
%
Assuming the contribution from the continuum states can be
approximated well by the spectral density of OPE above a threshold
value $s_0$ (duality), we arrive at the sum rule equation
%
%%%%%%%%%%%%%%%%%%%%%%%%%%%%%%%%%%%%%%%%%%%%%%%%%%%%%%%%%%%%%%%%%%%%%%%%%%%%%%
\begin{equation}
\Pi(M_B^2) \equiv f^2_Y e^{-M_Y^2/M_B^2} = \int^{s_0}_{s_<}
e^{-s/M_B^2}\rho(s)ds \label{eq:fin} \, .
\end{equation}
%%%%%%%%%%%%%%%%%%%%%%%%%%%%%%%%%%%%%%%%%%%%%%%%%%%%%%%%%%%%%%%%%%%%%%%%%%%%%%
%
Differentiating Eq.~(\ref{eq:fin}) with respect to $1 / M_B^2$ and
dividing it by Eq. (\ref{eq:fin}), finally we obtain
%
%%%%%%%%%%%%%%%%%%%%%%%%%%%%%%%%%%%%%%%%%%%%%%%%%%%%%%%%%%%%%%%%%%%%%%%%%%%%%%
\begin{equation}
M^2_Y =
\frac{\frac{\partial}{\partial(-1/M_B^2)}\Pi(M_B^2)}{\Pi(M_B^2)} =
\frac{\int^{s_0}_{s_<} e^{-s/M_B^2}s\rho(s)ds}{\int^{s_0}_{s_<}
e^{-s/M_B^2}\rho(s)ds}\, . \label{eq:LSR}
\end{equation}
%%%%%%%%%%%%%%%%%%%%%%%%%%%%%%%%%%%%%%%%%%%%%%%%%%%%%%%%%%%%%%%%%%%%%%%%%%%%%%
%
In the following, we study both Eqs.~(\ref{eq:fin}) and
(\ref{eq:LSR}) as functions of the parameters such as the Borel mass
$M_B$ and the threshold value $s_0$ for various combinations of the
tetraquark currents.

We have performed the OPE calculation up to dimension twelve. Here
we only show the results for currents $\eta^M_1$ and $\eta^M_5$,
which have quark contents $q q \bar q \bar q$ and $q s \bar q \bar
s$, respectively. Others are shown in the Appendix.~\ref{app:ope}.
%
%%%%%%%%%%%%%%%%%%%%%%%%%%%%%%%%%%%%%%%%%%%%%%%%%%%%%%%%%%%%%%%%%%%%%%%%%%%%%%
\begin{eqnarray}
%------------------------------\rho 1600 1----------------------------------
\Pi^M_1(M_B^2) &=& \int^{s_0}_{0} \Bigg [ {1 \over 18432 \pi^6} s^4
- { \langle g_s^2 G G \rangle \over 18432 \pi^6 }  s^2 + { \langle
\bar q q \rangle^2 \over 18 \pi^2 }  s + { \langle \bar q q \rangle
\langle g_s \bar q \sigma G q \rangle \over 12 \pi^2 } \Bigg ]
e^{-s/M_B^2} ds
\\ \nonumber && + \Big ( { \langle g_s \bar q \sigma G q \rangle^2 \over 48 \pi^2
} - { 5 \langle g_s^2 GG \rangle \langle \bar q q \rangle ^2 \over
864 \pi^2 }  \Big ) + {1 \over M_B^2}\Big( - {32 g_s^2 \langle \bar
q q \rangle ^4 \over 81 } + { \langle g_s^2 GG \rangle \langle \bar
q q \rangle \langle g_s \bar q \sigma G q \rangle \over 576 \pi^2 }
\Big)\, .
\end{eqnarray}
%%%%%%%%%%%%%%%%%%%%%%%%%%%%%%%%%%%%%%%%%%%%%%%%%%%%%%%%%%%%%%%%%%%%%%%%%%%%%%
%
%
%%%%%%%%%%%%%%%%%%%%%%%%%%%%%%%%%%%%%%%%%%%%%%%%%%%%%%%%%%%%%%%%%%%%%%%%%%%%%%
\begin{eqnarray}
%------------------------------\rho 2000 1----------------------------------
\Pi^M_5(M_B^2) &=& \int^{s_0}_{4 m_s^2} \Bigg [ {1 \over 18432
\pi^6} s^4 - { 17 m_s^2 \over 7680 \pi^6 } s^3 + \Big ( - { \langle
g_s^2 G G \rangle \over 18432 \pi^6 } - {m_s \langle \bar q q
\rangle \over 96 \pi^4} + {m_s \langle \bar s s \rangle \over
48\pi^4} \Big ) s^2 + \Big ( - { \langle \bar q q \rangle^2 \over 36
\pi^2 } + { \langle \bar q q \rangle \langle \bar s s \rangle \over
9 \pi^2 }
\\ \nonumber && - { \langle \bar s s \rangle^2 \over 36 \pi^2 } - { m_s
\langle g_s \bar q \sigma G q \rangle \over 48 \pi^4 } + { m_s
\langle g_s \bar s \sigma G s \rangle \over 96 \pi^4 } + { m_s^2
\langle g_s^2 G G \rangle \over 4608 \pi^6 } \Big ) s - { \langle
\bar q q \rangle \langle g_s \bar q \sigma G q \rangle \over 24
\pi^2 } + { \langle \bar q q \rangle \langle g_s \bar s \sigma G s
\rangle \over 12\pi^2 }
\\ \nonumber &&  + { \langle \bar s s \rangle \langle g_s \bar q \sigma G q \rangle
\over 12 \pi^2 } - { \langle \bar s s \rangle \langle g_s \bar s
\sigma G s \rangle \over 24 \pi^2 } + { m_s \langle g_s^2 G G
\rangle \langle \bar q q \rangle \over 256 \pi^4} - { m_s^2 \langle
\bar q q \rangle^2 \over 6 \pi^2 } - { m_s^2 \langle \bar q q
\rangle \langle \bar s s \rangle \over 2 \pi^2 } + { m_s^2 \langle
\bar s s \rangle^2 \over 24 \pi^2 } \Bigg ] e^{-s/M_B^2} ds
\\ \nonumber && + \Big ( - { \langle g_s \bar q \sigma G q \rangle^2 \over 96 \pi^2
} + { \langle g_s \bar q \sigma G q \rangle \langle g_s \bar s
\sigma G s \rangle \over 24 \pi^2 } - { \langle g_s \bar s \sigma G
s \rangle^2 \over 96 \pi^2 } - { 5 \langle g_s^2 GG \rangle \langle
\bar q q \rangle \langle \bar s s \rangle \over 864 \pi^2 } + { 2m_s
\langle \bar q q \rangle^2 \langle \bar s s \rangle \over 3 } + { 4
m_s \langle \bar q q \rangle \langle \bar s s \rangle^2 \over 9 }
\\
\nonumber &&  + { 5 m_s \langle g_s^2 GG \rangle \langle g_s \bar q
\sigma G q \rangle \over 4608 \pi^4 } - { m_s^2 \langle \bar s s
\rangle \langle g_s \bar q \sigma G q \rangle \over 4 \pi^2 } - {
m_s^2 \langle \bar q q \rangle \langle g_s \bar s \sigma G s \rangle
\over 6 \pi^2 } \Big ) + {1 \over M_B^2} \Big( - {32 g_s^2 \langle
\bar q q \rangle ^2 \langle \bar s s \rangle^2 \over 81 } \\
\nonumber && + { \langle g_s^2 GG \rangle \langle \bar q q \rangle
\langle g_s \bar s \sigma G s \rangle \over 1152 \pi^2 } + { \langle
g_s^2 GG \rangle \langle \bar s s \rangle \langle g_s \bar q \sigma
G q \rangle \over 1152 \pi^2 } - { 2 m_s \langle \bar q q \rangle^2
\langle g_s \bar s \sigma G s \rangle \over 9} - { 5 m_s \langle
\bar q q \rangle \langle \bar s s \rangle \langle g_s \bar q \sigma
G q \rangle \over 9}
\\ \nonumber && + { m_s \langle \bar q q \rangle \langle \bar s s \rangle \langle g_s
\bar s \sigma G s \rangle \over 9} + { m_s \langle \bar s s
\rangle^2 \langle g_s \bar q \sigma G q \rangle \over 9} + { m_s^2
\langle g_s \bar q \sigma G q \rangle^2 \over 24 \pi^2 } - { m_s^2
\langle g_s \bar q \sigma G q \rangle \langle g_s \bar s \sigma G s
\rangle \over 24 \pi^2 } \Big)\, .
\end{eqnarray}
%%%%%%%%%%%%%%%%%%%%%%%%%%%%%%%%%%%%%%%%%%%%%%%%%%%%%%%%%%%%%%%%%%%%%%%%%%%%%%
%
In the above equations, $\langle \bar{s}s \rangle$ is the dimension
$D=3$ strange quark condensate; $\langle g^2 GG \rangle$ is a $D=4$
gluon condensate; $\langle g\bar{s}\sigma Gs \rangle$ is $D=5$ mixed
condensate. There are many terms which give minor contributions,
such as $\langle g^3 G^3 \rangle$, and we omit them. As usual, we
assume the vacuum saturation for higher dimensional condensates such
as $\langle 0 | \bar q q \bar q q |0\rangle \sim \langle 0 | \bar q
q |0\rangle \langle 0|\bar q q |0\rangle$. To obtain these results,
we keep the terms of order $O(m_q^2)$ in the propagators of a
massive quark in the presence of quark and gluon condensates:
%
%%%%%%%%%%%%%%%%%%%%%%%%%%%%%%%%%%%%%%%%%%%%%%%%%%%%%%%%%%%%%%%%%%%%%%%%%%%%%%
\begin{eqnarray} \nonumber
i S^{ab} & \equiv & \langle 0 | T [ q^a(x) q^b(0) ] | 0 \rangle
\\ \nonumber &=& { i \delta^{ab} \over 2 \pi^2 x^4 } \hat{x} + {i \over
32\pi^2} { \lambda^n_{ab} \over 2 } g_c G^n_{\mu\nu} {1 \over x^2}
(\sigma^{\mu\nu} \hat{x} + \hat{x} \sigma^{\mu\nu}) - { \delta^{ab}
\over 12 } \langle \bar q q \rangle \\ && + { \delta^{ab} x^2 \over
192 } \langle g_c \bar q \sigma G q \rangle - { m_q \delta^{ab}
\over 4 \pi^2 x^2 } + { i \delta^{ab} m_q \langle \bar q q \rangle
\over 48 }  \hat x + { i \delta^{ab} m_q^2 \over 8 \pi^2 x^2 }
\hat{x} \, .
\end{eqnarray}
%%%%%%%%%%%%%%%%%%%%%%%%%%%%%%%%%%%%%%%%%%%%%%%%%%%%%%%%%%%%%%%%%%%%%%%%%%%%%%
%

%
%=====================================================================================
%=====================================================================================
\section{Numerical Analysis}\label{sec:numeri}
%=====================================================================================
%=====================================================================================
%

In our numerical analysis, we use the following values for various
condensates and $m_s$ at 1 GeV and $\alpha_s$ at 1.7 GeV
~\cite{Yang:1993bp,Narison:2002pw,Gimenez:2005nt,Jamin:2002ev,Ioffe:2002be,Ovchinnikov:1988gk,Yao:2006px}:
%
%%%%%%%%%%%%%%%%%%%%%%%%%%%%%%%%%%%%%%%%%%%%%%%%%%%%%%%%%%%%%%%%%%%%%%%%%%%%%%
\begin{eqnarray}
\nonumber &&\langle\bar qq \rangle=-(0.240 \mbox{ GeV})^3\, ,
\\
\nonumber &&\langle\bar ss\rangle=-(0.8\pm 0.1)\times(0.240 \mbox{
GeV})^3\, ,
\\
\nonumber &&\langle g_s^2GG\rangle =(0.48\pm 0.14) \mbox{ GeV}^4\, ,
\\
\label{condensates} && \langle g_s\bar q\sigma G
q\rangle=-M_0^2\times\langle\bar qq\rangle\, ,
\\
\nonumber && M_0^2=(0.8\pm0.2)\mbox{ GeV}^2\, ,
\\
\nonumber &&m_s(1\mbox{ GeV})=125 \pm 20 \mbox{ MeV}\, ,
\\
\nonumber && \alpha_s(1.7\mbox{GeV}) = 0.328 \pm 0.03 \pm 0.025 \, .
\end{eqnarray}
%%%%%%%%%%%%%%%%%%%%%%%%%%%%%%%%%%%%%%%%%%%%%%%%%%%%%%%%%%%%%%%%%%%%%%%%%%%%%%
%
There is a minus sign in the definition of the mixed condensate
$\langle g_s\bar q\sigma G q\rangle$, which is different from that
used in some other QCD sum rule studies. This difference just comes
from the definition of coupling constant
$g_s$~\cite{Yang:1993bp,Hwang:1994vp}.

\begin{figure}[h!t]
\begin{center}
\scalebox{0.75}{\includegraphics{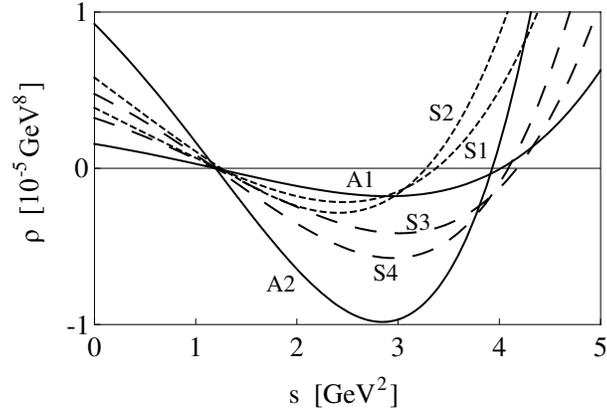}} \caption{Spectral
densities for the current $\eta^A_{1\mu}$, $\eta^A_{2\mu}$ (solid
lines), $\eta^S_{1\mu}$, $\eta^S_{2\mu}$ (short-dashed lines),
$\eta^S_{3\mu}$ and $\eta^S_{4\mu}$ (long-dashed lines). The labels
besides the lines indicate the flavor symmetry ($S$ or $A$) and
suffix i of the current $\eta^{S,A}_{i\mu}$ ($i=1,2,3,4$).}
\label{fig:specAS}
\end{center}
\end{figure}

\begin{figure}[h!t]
\begin{center}
\scalebox{0.75}{\includegraphics{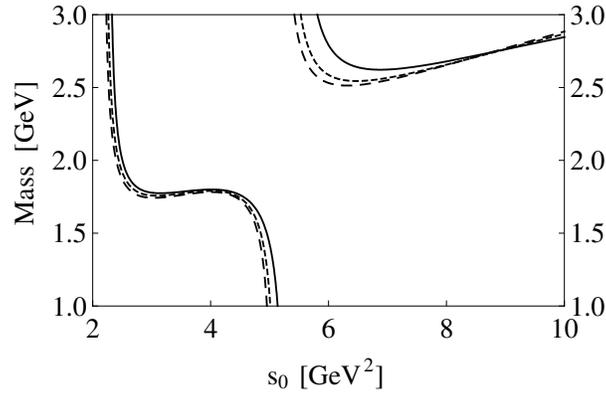}} \caption{The mass
calculated by using the current $\eta^A_{1\mu}$, as functions of
$s_0$ in units of GeV. The curves are obtained by setting $M_B^2 =
2$ GeV$^2$ (solid line), 3 GeV$^2$ (short-dashed line) and 4 GeV$^2$
(long-dashed line). The left curves (disconnected from the right
part) are obtained from a negative correlation function, and have no
physical meaning.} \label{fig:etaA1}
\end{center}
\end{figure}

For the currents which belong to the flavor representation
$\mathbf{6_f} \otimes \mathbf{ \bar 6_f }$ ($\mathbf{S}$) and
$\mathbf{ \bar 3_f} \otimes \mathbf{ 3_f }$ ($\mathbf{A}$), the
spectral densities turn out to be negative in the energy region $1$
GeV $\sim$ $2$ GeV as shown in Fig.~\ref{fig:specAS}. The spectral
densities of these currents become positive in the region $s > 4$
GeV$^2$. They may couple to the state $\pi_1(2015)$. However, after
performing the sum rule calculation, we find that the mass obtained
from the currents $\eta^A_{i\mu}$ and $\eta^S_{i\mu}$ is larger than
2.5 GeV, for instance, we show the mass calculated from the current
$\eta^A_{1\mu}$ in Fig.~\ref{fig:etaA1}. The curves are obtained by
setting $M_B^2 = 2$ GeV$^2$ (solid line), 3 GeV$^2$ (short-dashed
line) and 4 GeV$^2$ (long-dashed line). The left curves
(disconnected from the right part) are obtained from a negative
Borel transformed correlation function, and have no physical
meaning. Therefore, our QCD sum rule analysis does not support
$\pi_1(1400)$, $\pi_1(1600)$ and $\pi_1(2015)$ as tetraquark states
with a flavor structure either $\mathbf{6_f} \otimes \mathbf{ \bar
6_f }$ or $\mathbf{\bar 3_f} \otimes \mathbf{ 3_f }$.

%
%%%%%%%%%%%%%%%%%%%%%%%%%%%%%%%%%%%%%%%%%%%%%%%%%%%%%%%%%%%%%%%%%%%%%%%%%%%%%%
%---------figure 6*6
%%%%%%%%%%%%%%%%%%%%%%%%%%%%%%%%%%%%%%%%%%%%%%%%%%%%%%%%%%%%%%%%%%%%%%%%%%%%%%
%

%
%%%%%%%%%%%%%%%%%%%%%%%%%%%%%%%%%%%%%%%%%%%%%%%%%%%%%%%%%%%%%%%%%%%%%%%%%%%%%%
%---------figure 6*6
\begin{figure}[h!t]
\begin{center}
\scalebox{0.75}{\includegraphics{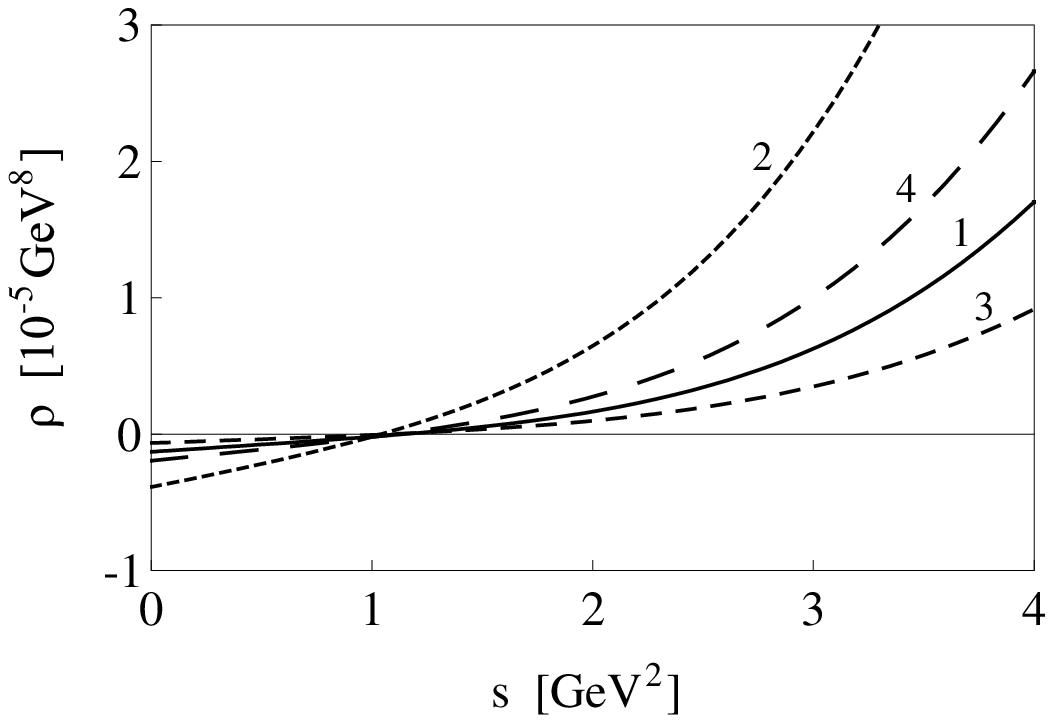}}
\scalebox{0.75}{\includegraphics{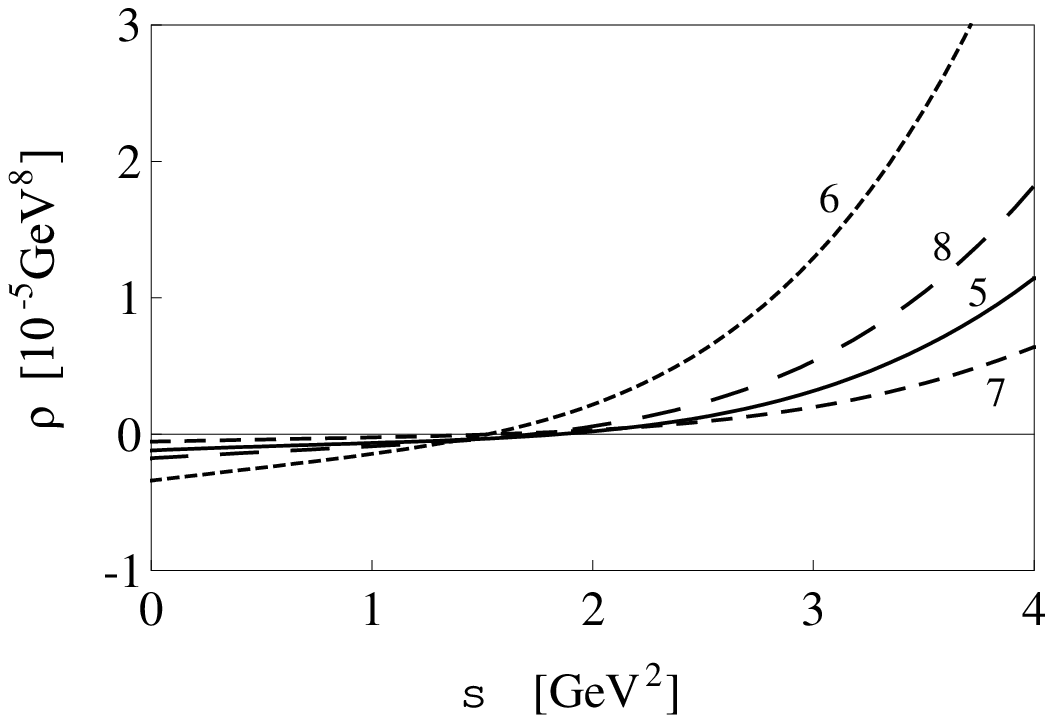}}\caption{Spectral
densities for the current $\eta^M_{i\mu}$. The spectral densities
for the currents with the quark contents $q q \bar q \bar q$ are
shown in the left hand side, and those with the quark contents $q s
\bar q \bar s$ are shown in the right hand side. The labels besides
the lines indicate the suffix i of the current $\eta^M_{i\mu}$
($i=1,\cdots,8$).} \label{fig:specM}
\end{center}
\end{figure}
%%%%%%%%%%%%%%%%%%%%%%%%%%%%%%%%%%%%%%%%%%%%%%%%%%%%%%%%%%%%%%%%%%%%%%%%%%%%%%
%

%
%%%%%%%%%%%%%%%%%%%%%%%%%%%%%%%%%%%%%%%%%%%%%%%%%%%%%%%%%%%%%%%%%%%%%%%%%%%%%%
%---------figure 6*6
%%%%%%%%%%%%%%%%%%%%%%%%%%%%%%%%%%%%%%%%%%%%%%%%%%%%%%%%%%%%%%%%%%%%%%%%%%%%%%
%

\begin{figure}[h!t]
\begin{center}
\scalebox{0.7}{\includegraphics{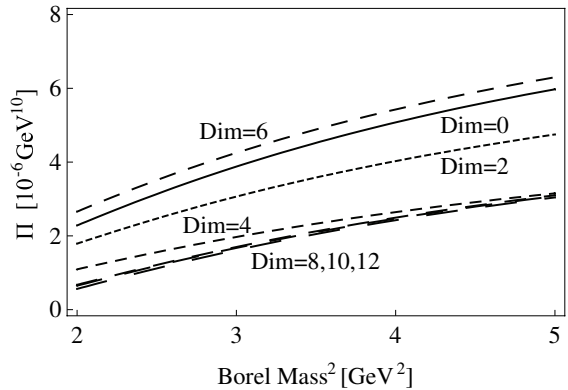}} \caption{Various
contribution to the correlation function for the current
$\eta^M_{5\mu}$ as functions of the Borel mass $M_B$ in units of
GeV$^10$ at $s_0$ = 4 GeV$^2$. The labels indicate the dimension up
to which the OPE terms are included.} \label{fig:pi}
\end{center}
\end{figure}

When using the currents $\eta^M_{i\mu}$, the spectral densities are
positive as shown in Fig.~\ref{fig:specM}. And so we shall use these
currents to perform a QCD sum rule analysis. First we need to study
the convergence of the OPE. The Borel transformed correlation
function of the current $\eta^M_{5\mu}$ is shown in
Fig.~\ref{fig:pi}, when we take $s_0 = 4$ GeV$^2$. Besides the first
term, which is the continuum piece, the D=6 and D=8 terms give large
contributions. The D=6 terms contain $\langle \bar q q \rangle^2$
and the D=8 terms contain $\langle \bar q q \rangle \langle g_c \bar
q \sigma G q \rangle$, which are the important condensates. We find
that the convergence is very good in the region of
$2$~GeV$^2<M_B^2<$~5GeV$^2$. Therefore, in this region, OPEs are
reliable.
%
%%%%%%%%%%%%%%%%%%%%%%%%%%%%%%%%%%%%%%%%%%%%%%%%%%%%%%%%%%%%%%%%%%%%%%%%%%%%%%
%---------figure 6*6
%%%%%%%%%%%%%%%%%%%%%%%%%%%%%%%%%%%%%%%%%%%%%%%%%%%%%%%%%%%%%%%%%%%%%%%%%%%%%%
%

\begin{figure}[h!t]
\begin{center}
\scalebox{0.7}{\includegraphics{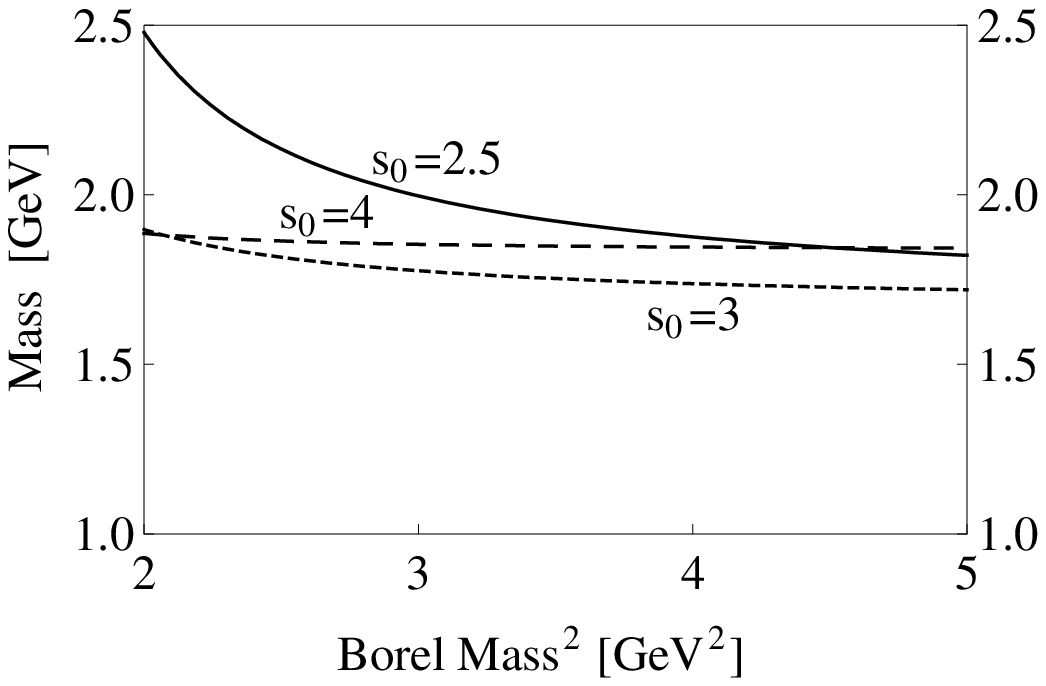}}
\scalebox{0.7}{\includegraphics{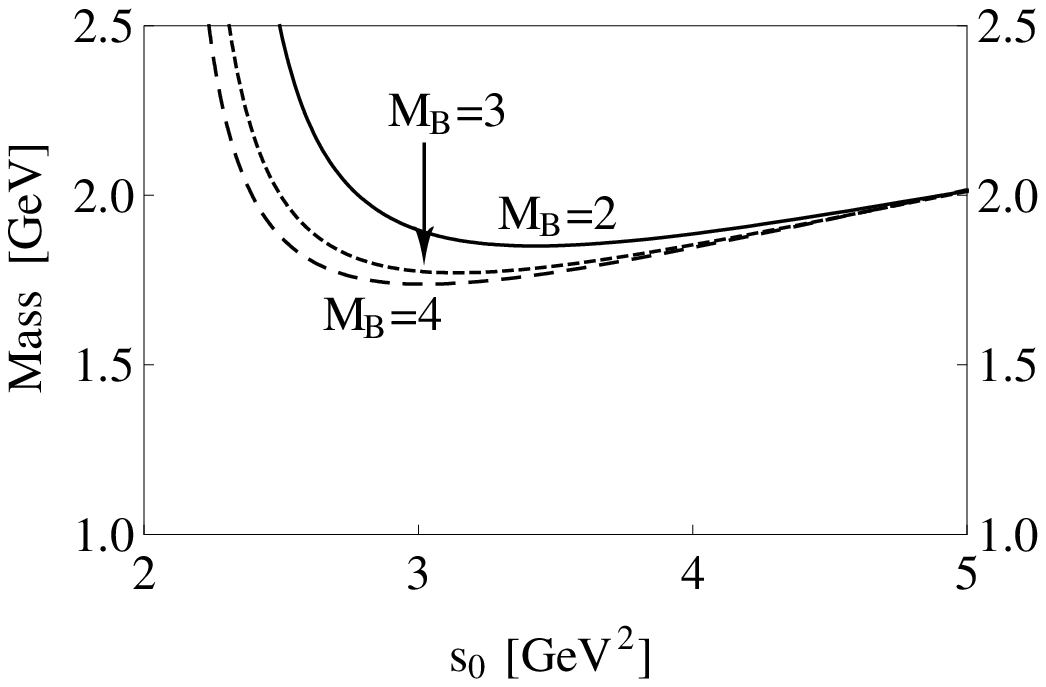}} \caption{The
mass of the state $q q \bar q \bar q$ calculated by using the
current $\eta^M_{1\mu}$, as functions of $M_B^2$ (Left) and $s_0$
(Right) in units of GeV.}\label{fig:eta1}
\end{center}
\end{figure}

\begin{figure}[h!t]
\begin{center}
\scalebox{0.7}{\includegraphics{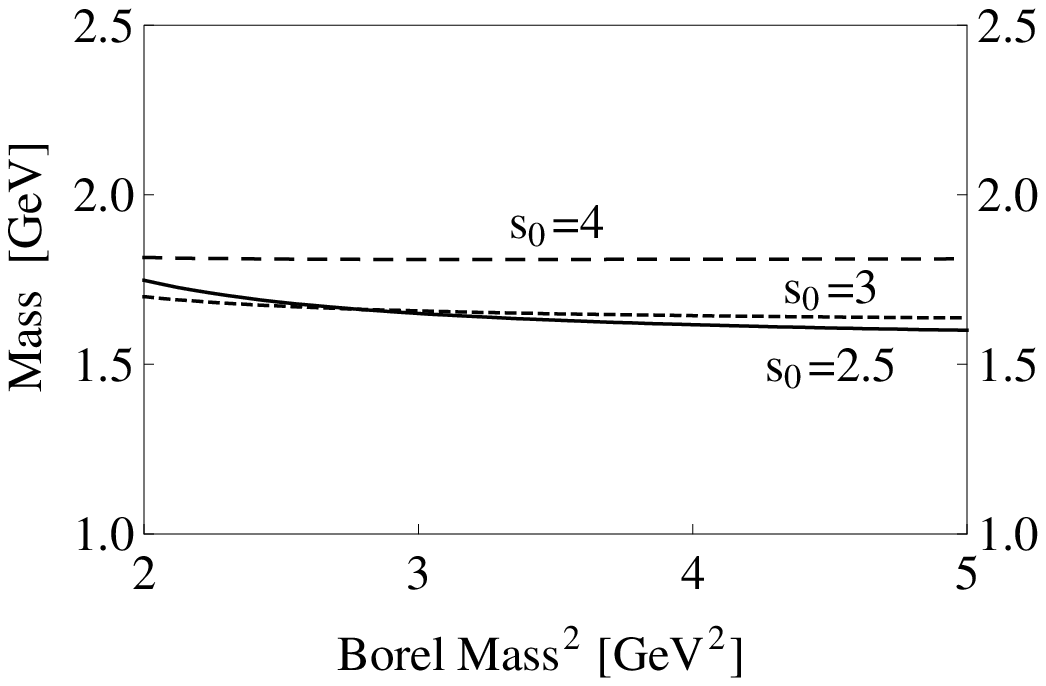}}
\scalebox{0.7}{\includegraphics{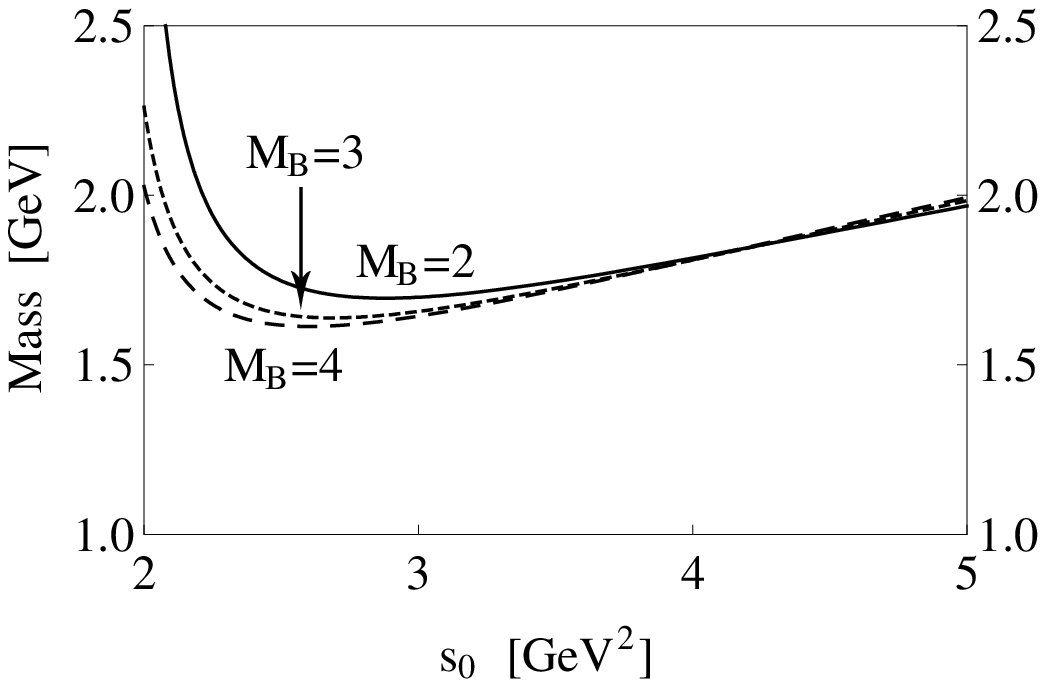}} \caption{The
mass of the state $q q \bar q \bar q$ calculated by using the
current $\eta^M_{2\mu}$, as functions of $M_B^2$ (Left) and $s_0$
(Right) in units of GeV.}\label{fig:eta2}
\end{center}
\end{figure}

\begin{figure}[!t]
\begin{center}
\scalebox{0.7}{\includegraphics{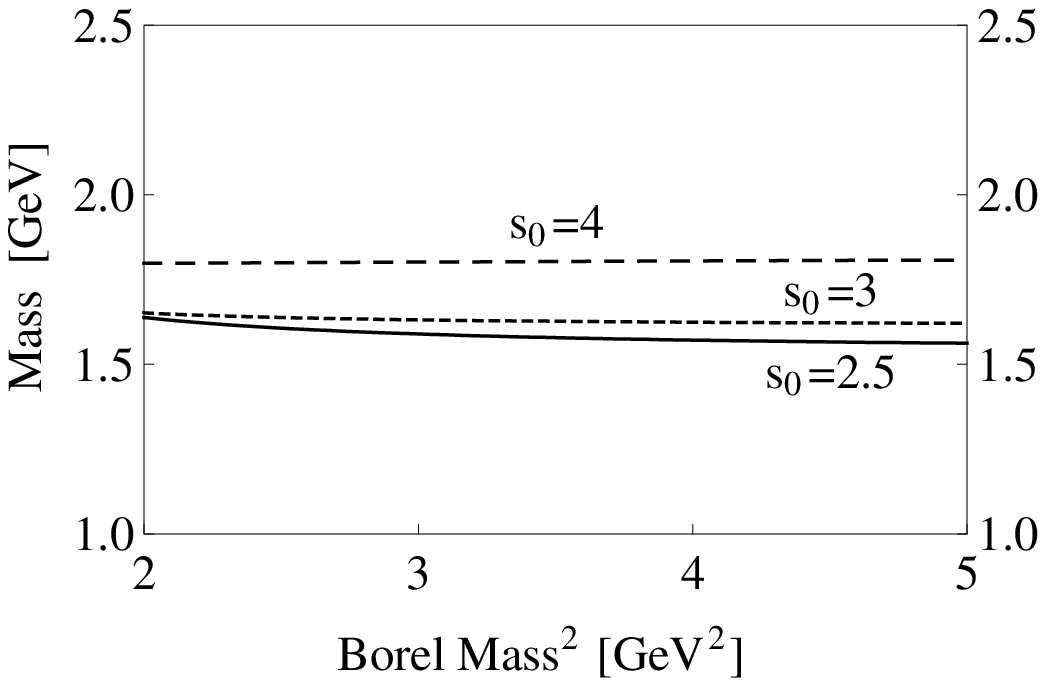}}
\scalebox{0.7}{\includegraphics{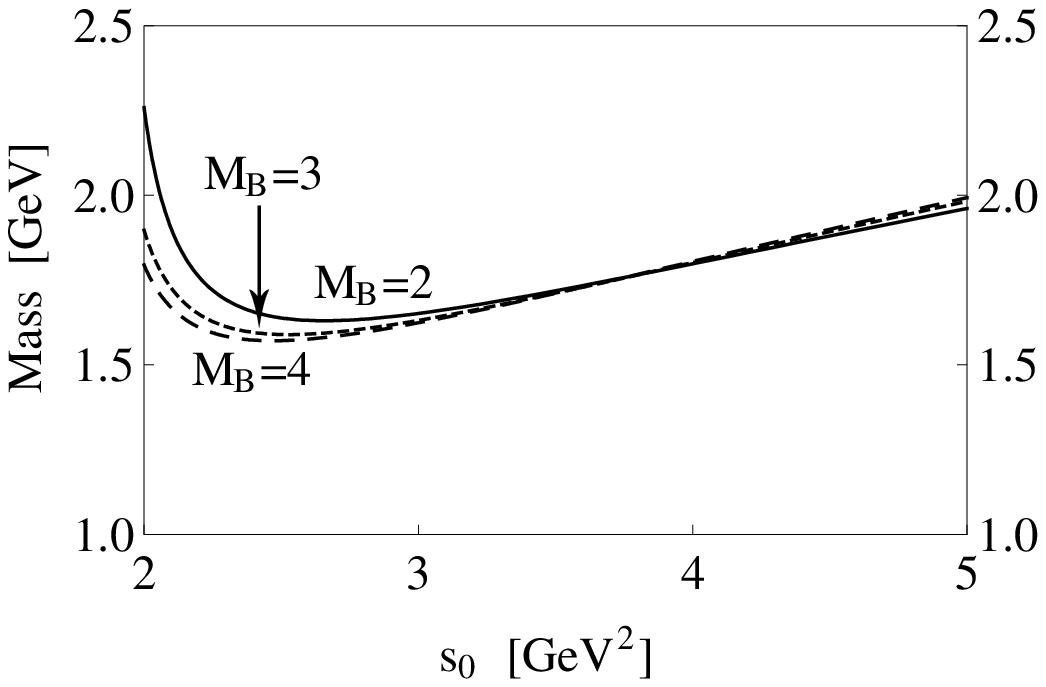}} \caption{The
mass of the state $q q \bar q \bar q$ calculated by using the
current $\eta^M_{3\mu}$, as functions of $M_B^2$ (Left) and $s_0$
(Right) in units of GeV.}\label{fig:eta3}
\end{center}
\end{figure}

\begin{figure}[!t]
\begin{center}
\scalebox{0.7}{\includegraphics{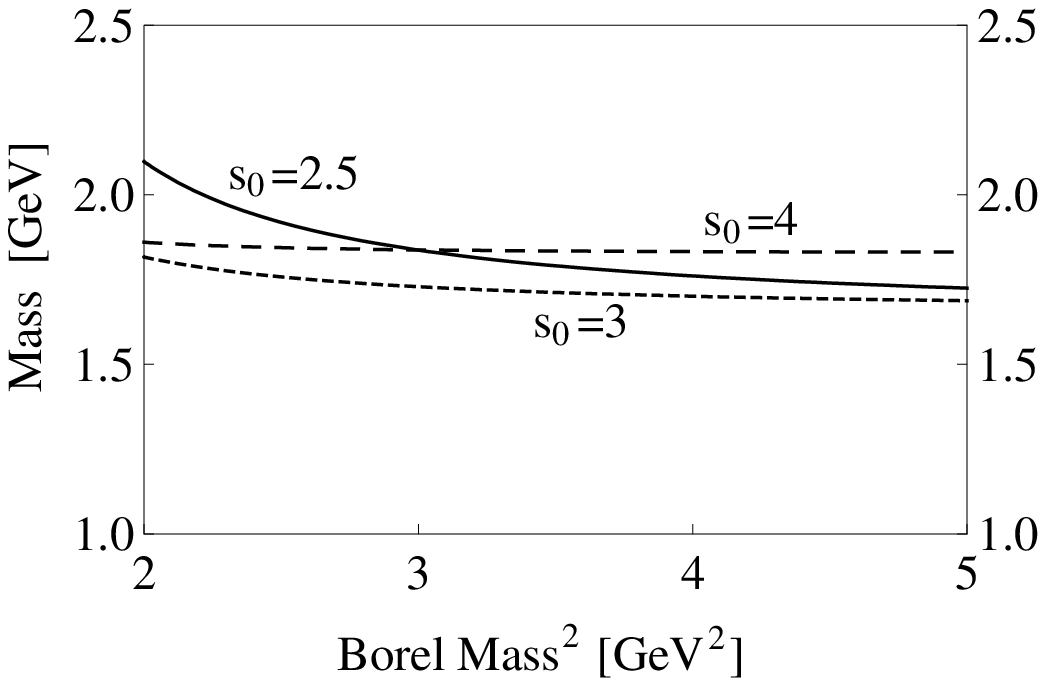}}
\scalebox{0.7}{\includegraphics{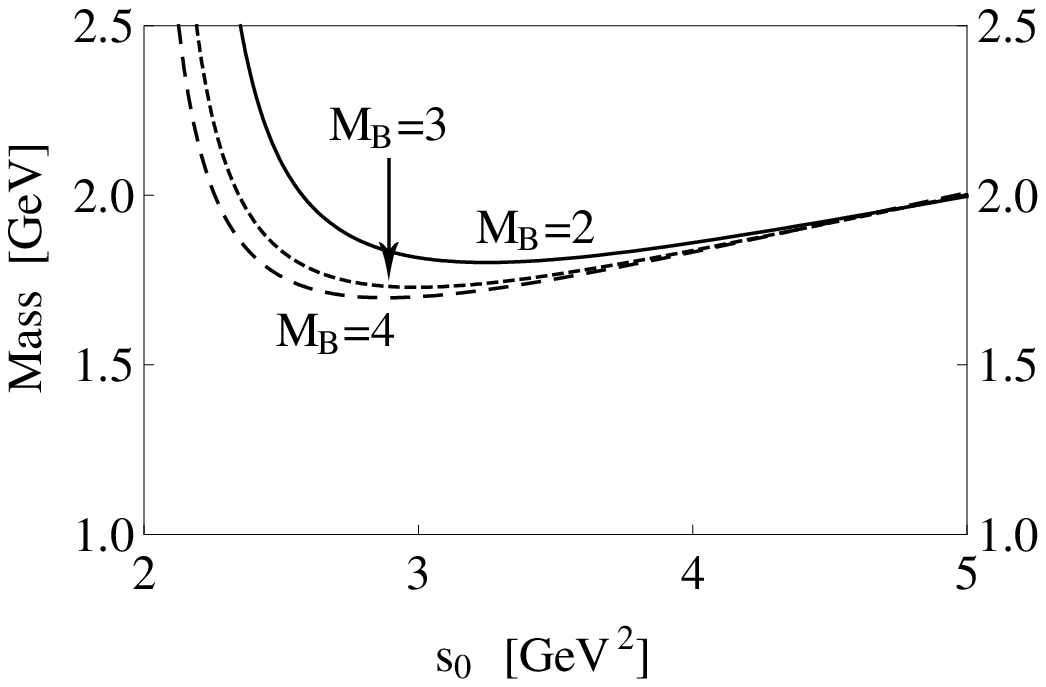}} \caption{The
mass of the state $q q \bar q \bar q$ calculated by using the
current $\eta^M_{4\mu}$, as functions of $M_B^2$ (Left) and $s_0$
(Right) in units of GeV.}\label{fig:eta4}
\end{center}
\end{figure}

The mass is calculated by using Eq.~(\ref{eq:LSR}), and results are
obtained as functions of Borel mass $M_B$ and threshold value $s_0$.
In Figs.~\ref{fig:eta1}, \ref{fig:eta2}, \ref{fig:eta3} and
\ref{fig:eta4}, we show the mass calculated from currents
$\eta^M_{1\mu}$, $\eta^M_{2\mu}$, $\eta^M_{3\mu}$ and
$\eta^M_{4\mu}$, whose quark contents are $q q \bar q \bar q$.
Although these four independent currents look much different, we
find that they give a similar result. From figures at LHS, we find
that the dependence on Borel mass is weak. From figures at RHS where
the mass is shown as functions of $s_0$, we find that there is a
mass minimum for all curves where the stability is the best. It is
1.7 GeV, 1.6 GeV, 1.6 GeV and 1.7 GeV for four independent currents,
respectively. We find that sometimes the threshold values become
smaller than the mass obtained in the mass minimum region. This is
due to the negative part of the spectral densities. We also met this
in the study of $Y(2175)$. See Ref~\cite{Chen:2008ej} for details.

%
%%%%%%%%%%%%%%%%%%%%%%%%%%%%%%%%%%%%%%%%%%%%%%%%%%%%%%%%%%%%%%%%%%%%%%%%%%%%%%
%%%%%%%%%%%%%%%%%%%%%%%%%%%%%%%%%%%%%%%%%%%%%%%%%%%%%%%%%%%%%%%%%%%%%%%%%%%%%%
%

%
%%%%%%%%%%%%%%%%%%%%%%%%%%%%%%%%%%%%%%%%%%%%%%%%%%%%%%%%%%%%%%%%%%%%%%%%%%%%%%
%%%%%%%%%%%%%%%%%%%%%%%%%%%%%%%%%%%%%%%%%%%%%%%%%%%%%%%%%%%%%%%%%%%%%%%%%%%%%%
%

%
%%%%%%%%%%%%%%%%%%%%%%%%%%%%%%%%%%%%%%%%%%%%%%%%%%%%%%%%%%%%%%%%%%%%%%%%%%%%%%
%%%%%%%%%%%%%%%%%%%%%%%%%%%%%%%%%%%%%%%%%%%%%%%%%%%%%%%%%%%%%%%%%%%%%%%%%%%%%%
%

%
%%%%%%%%%%%%%%%%%%%%%%%%%%%%%%%%%%%%%%%%%%%%%%%%%%%%%%%%%%%%%%%%%%%%%%%%%%%%%%
%%%%%%%%%%%%%%%%%%%%%%%%%%%%%%%%%%%%%%%%%%%%%%%%%%%%%%%%%%%%%%%%%%%%%%%%%%%%%%
%

\begin{figure}[!t]
\begin{center}
\scalebox{0.7}{\includegraphics{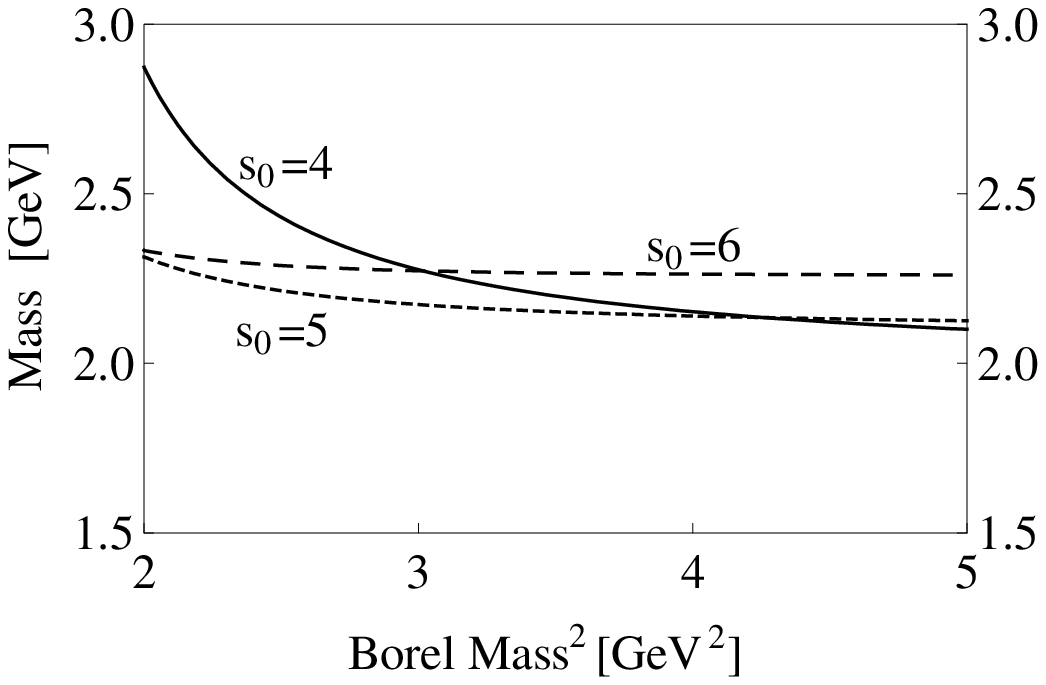}}
\scalebox{0.7}{\includegraphics{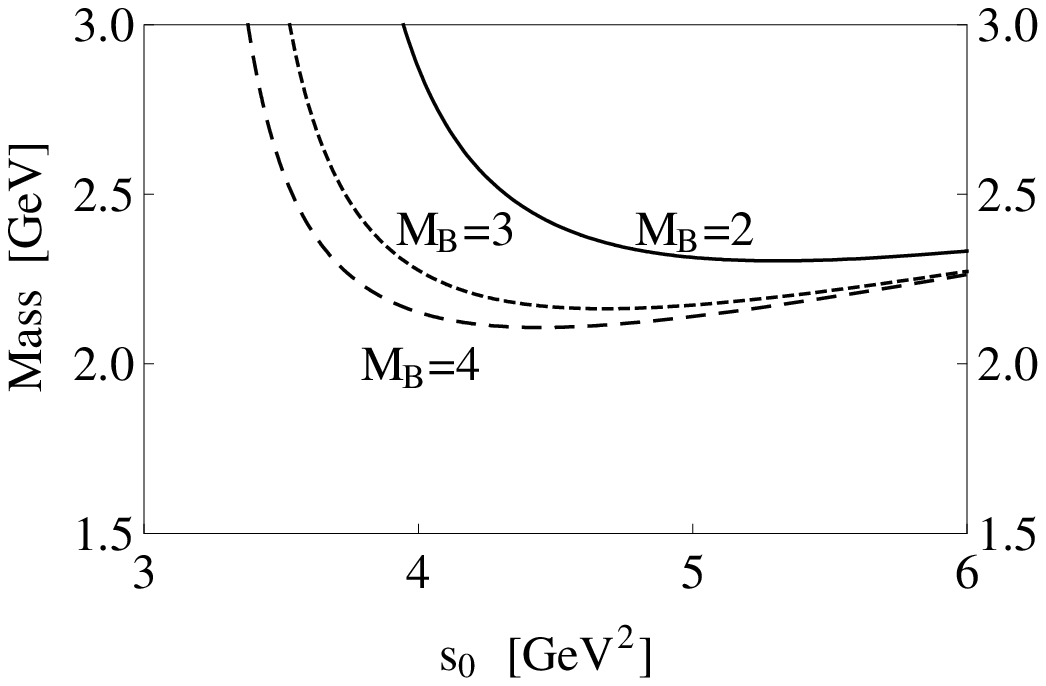}} \caption{The
mass of the state $q s \bar q \bar s$ calculated by using the
current $\eta^M_{5\mu}$, as functions of $M_B^2$ (Left) and $s_0$
(Right) in units of GeV.}\label{fig:eta5}
\end{center}
\end{figure}

\begin{figure}[!t]
\begin{center}
\scalebox{0.7}{\includegraphics{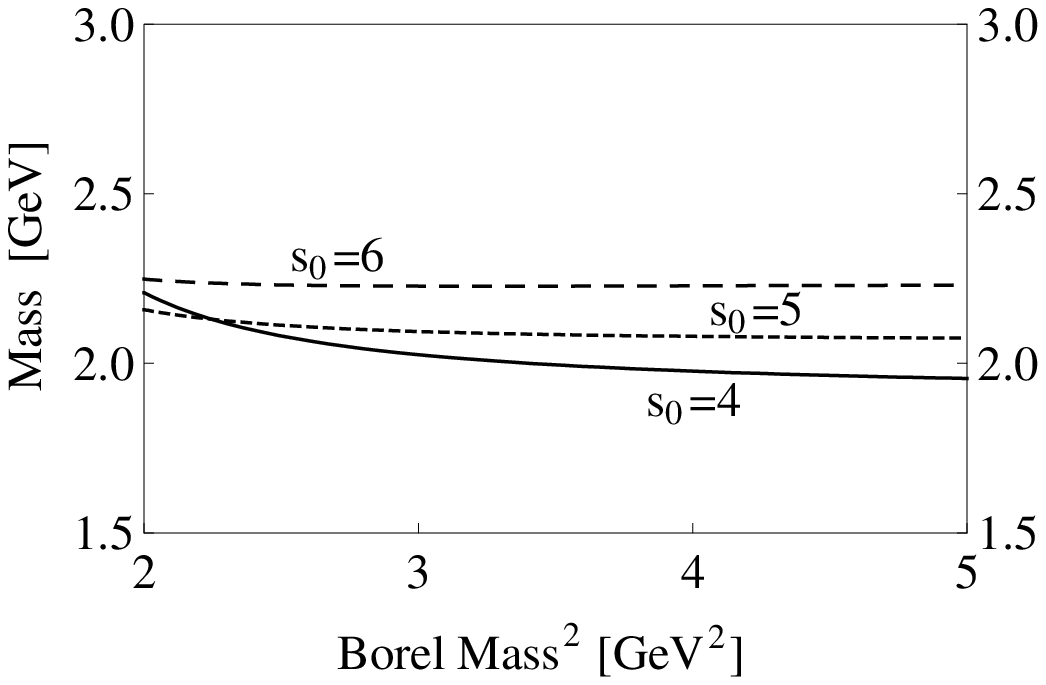}}
\scalebox{0.7}{\includegraphics{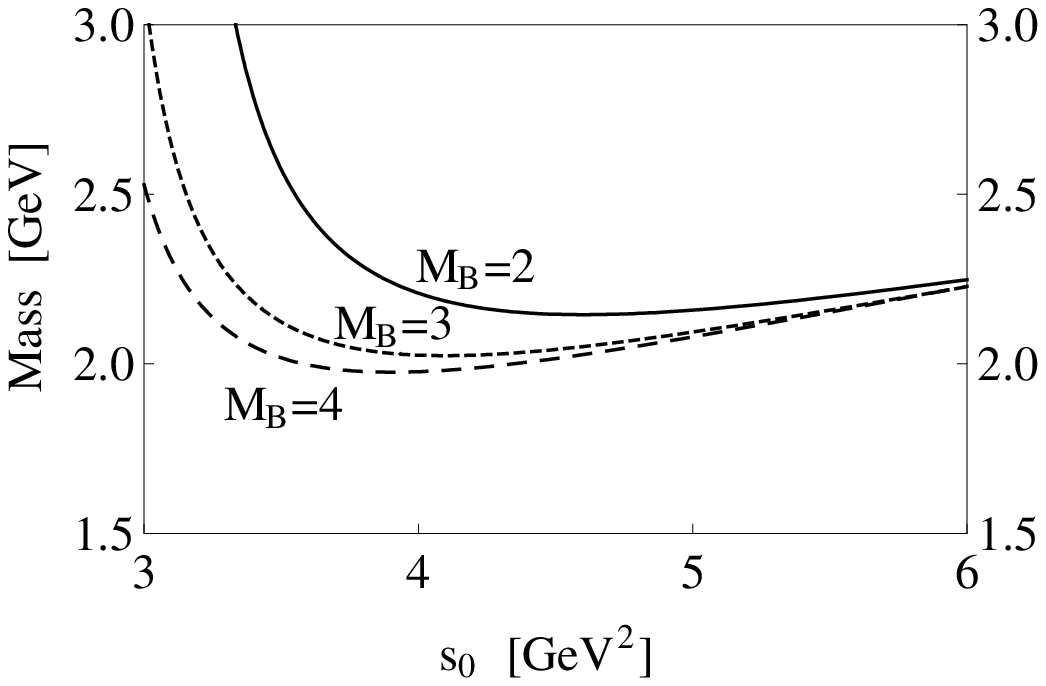}} \caption{The
mass of the state $q s \bar q \bar s$ calculated by using the
current $\eta^M_{6\mu}$, as functions of $M_B^2$ (Left) and $s_0$
(Right) in units of GeV.}\label{fig:eta6}
\end{center}
\end{figure}

\begin{figure}[!t]
\begin{center}
\scalebox{0.7}{\includegraphics{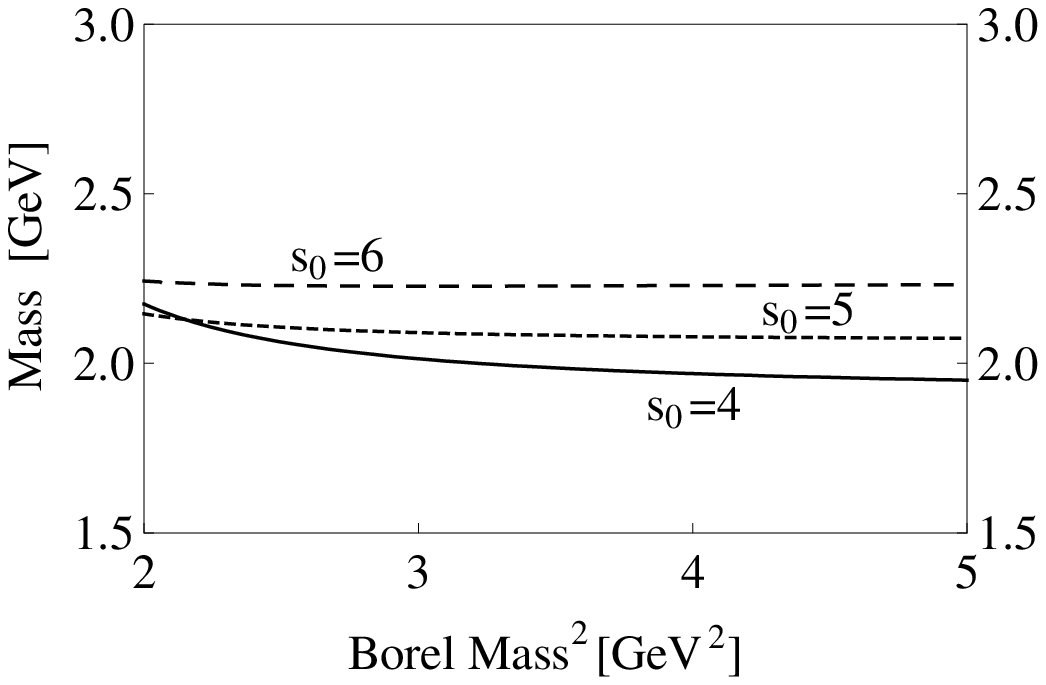}}
\scalebox{0.7}{\includegraphics{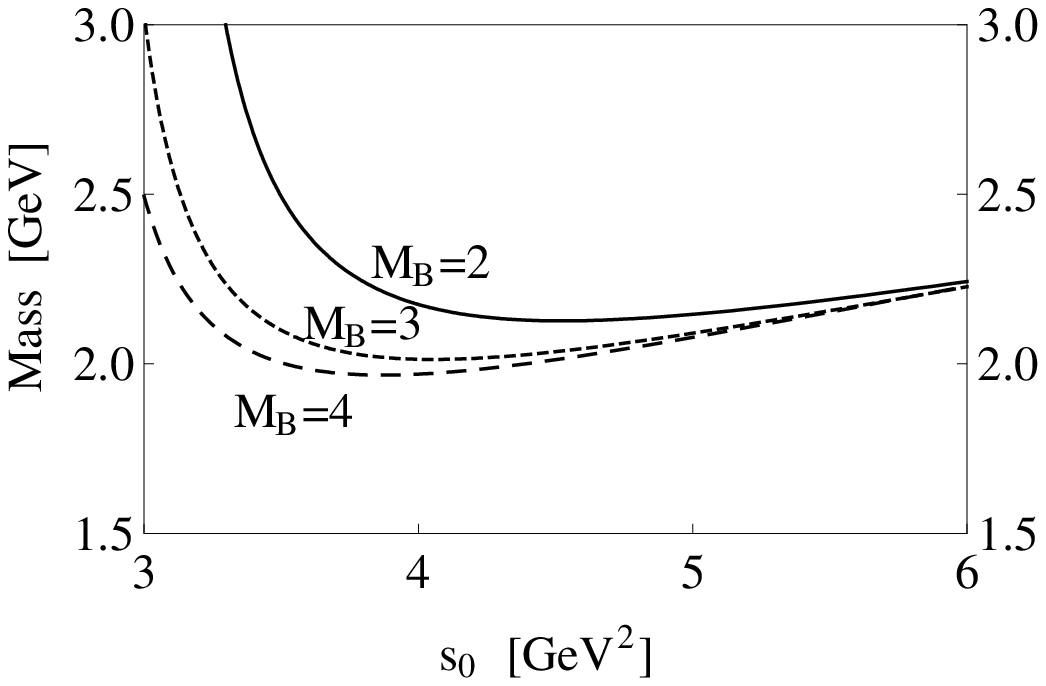}} \caption{The
mass of the state $q s \bar q \bar s$ calculated by using the
current $\eta^M_{7\mu}$, as functions of $M_B^2$ (Left) and $s_0$
(Right) in units of GeV.}\label{fig:eta7}
\end{center}
\end{figure}

\begin{figure}[!t]
\begin{center}
\scalebox{0.7}{\includegraphics{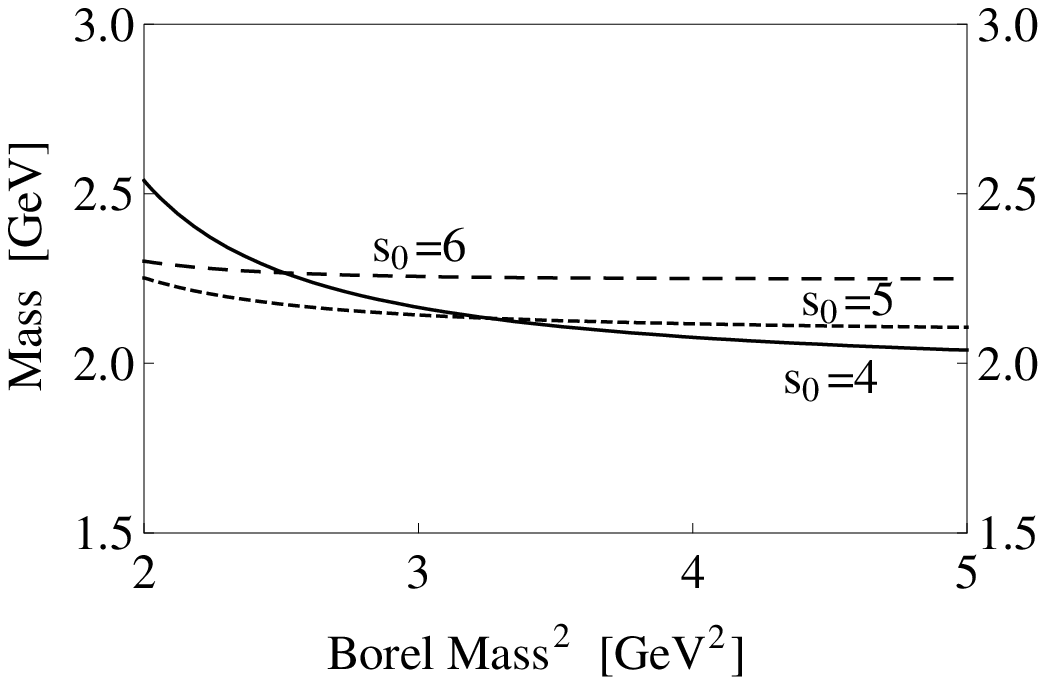}}
\scalebox{0.7}{\includegraphics{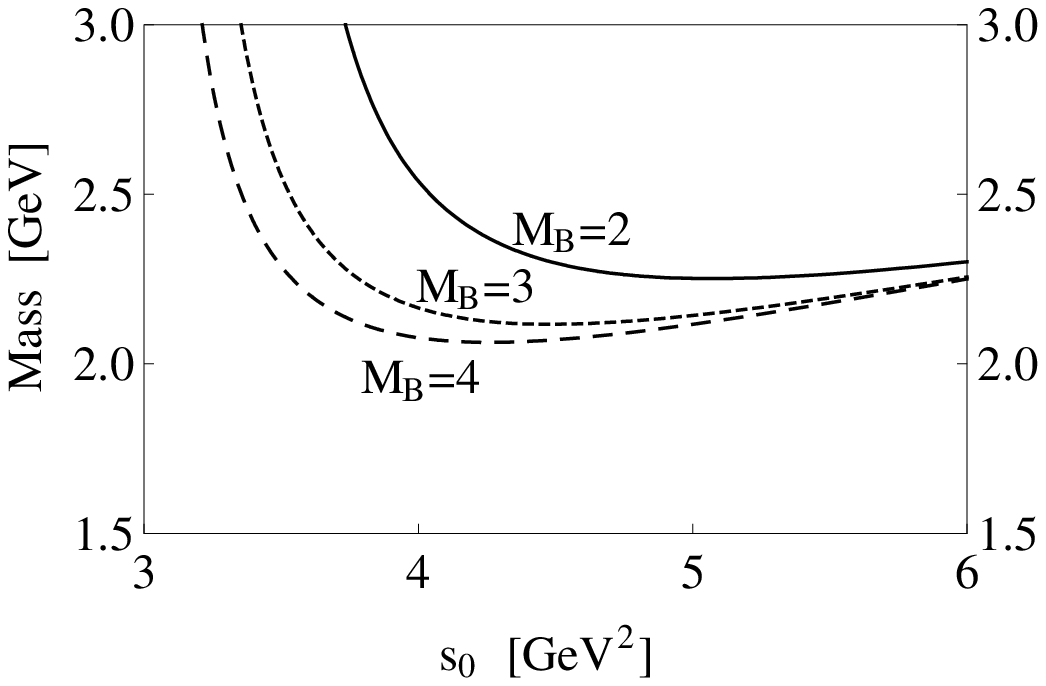}} \caption{The
mass of the state $q s \bar q \bar s$ calculated by using the
current $\eta^M_{8\mu}$, as functions of $M_B^2$ (Left) and $s_0$
(Right) in units of GeV.}\label{fig:eta8}
\end{center}
\end{figure}

In Figs.~\ref{fig:eta5}, \ref{fig:eta6}, \ref{fig:eta7} and
\ref{fig:eta8}, we show the mass calculated from currents
$\eta^M_{5\mu}$, $\eta^M_{6\mu}$, $\eta^M_{7\mu}$ and
$\eta^M_{8\mu}$, whose quark contents are $q s \bar q \bar s$. The
results are similar as previous four currents. But now the mass
obtained is about 0.4 GeV larger than the previous ones. The minimum
occurs at 2.1 GeV, 2.0 GeV, 1.9 GeV and 2.0 GeV, respectively.

%
%%%%%%%%%%%%%%%%%%%%%%%%%%%%%%%%%%%%%%%%%%%%%%%%%%%%%%%%%%%%%%%%%%%%%%%%%%%%%%
%%%%%%%%%%%%%%%%%%%%%%%%%%%%%%%%%%%%%%%%%%%%%%%%%%%%%%%%%%%%%%%%%%%%%%%%%%%%%%
%

%
%%%%%%%%%%%%%%%%%%%%%%%%%%%%%%%%%%%%%%%%%%%%%%%%%%%%%%%%%%%%%%%%%%%%%%%%%%%%%%
%%%%%%%%%%%%%%%%%%%%%%%%%%%%%%%%%%%%%%%%%%%%%%%%%%%%%%%%%%%%%%%%%%%%%%%%%%%%%%
%

%
%%%%%%%%%%%%%%%%%%%%%%%%%%%%%%%%%%%%%%%%%%%%%%%%%%%%%%%%%%%%%%%%%%%%%%%%%%%%%%
%%%%%%%%%%%%%%%%%%%%%%%%%%%%%%%%%%%%%%%%%%%%%%%%%%%%%%%%%%%%%%%%%%%%%%%%%%%%%%
%

%
%%%%%%%%%%%%%%%%%%%%%%%%%%%%%%%%%%%%%%%%%%%%%%%%%%%%%%%%%%%%%%%%%%%%%%%%%%%%%%
%%%%%%%%%%%%%%%%%%%%%%%%%%%%%%%%%%%%%%%%%%%%%%%%%%%%%%%%%%%%%%%%%%%%%%%%%%%%%%
%

In a short summary, we have performed a QCD sum rule analysis for $q
q \bar q \bar q$ and $q s \bar q \bar s$. The mass obtained is
around 1.6 GeV and 2.0 GeV, respectively. There are four independent
currents for each case, which give a similar results. Their mixing
would lead to a similar result, too. Compared with the experimental
data, they can be used to interpret the states $\pi_1(1600)$ and
$\pi_1(2015)$ of $I^GJ^{PC} = 1^-1^{-+}$. These analyses are very
similar to our previous paper~\cite{Chen:2008ej}, where we studied
the state $Y(2175)$ by using vector tetraquark currents which have
quantum numbers $J^{PC} = 1^{--}$ and quark contents $s s \bar s
\bar s$.

The pole contribution
\begin{equation}
{ \int^{s_0}_{s_<} e^{-s/M_B^2}\rho(s)ds \over \int^{\infty}_{s_<}
e^{-s/M_B^2}\rho(s)ds } \label{def:poleLSR}
\end{equation}
is not large enough for all currents due to the high dimension
nature of tetraquark currents. Another reason is that these currents
have a large coupling to the continuum, which is difficult to be
removed. Therefore, we arrive at a stable mass, but with a small
pole. To make our analysis more reliable, we go on to use the finite
energy sum rule.

%
%=====================================================================================
%=====================================================================================
\section{Finite Energy Sum Rule}\label{sec:fesr}
%=====================================================================================
%=====================================================================================
%

In this section, we use the method of finite energy sum rule (FESR).
In order to calculate the mass in the FESR, we first define the
$n$th moment by using the spectral function $\rho(s)$ in
Eq.~(\ref{eq:rho})
%
%%%%%%%%%%%%%%%%%%%%%%%%%%%%%%%%%%%%%%%%%%%%%%%%%%%%%%%%%%%%%%%%%%%%%%%%%%%%%%
\begin{equation}
W(n, s_0) = \int^{s_0}_0 \rho(s) s^n ds \, . \label{eq:moment}
\end{equation}
%%%%%%%%%%%%%%%%%%%%%%%%%%%%%%%%%%%%%%%%%%%%%%%%%%%%%%%%%%%%%%%%%%%%%%%%%%%%%%
%
This integral is used for the phenomenological side, while the
integral along the circular contour of radius $s_0$ on the $q^2$
complex plain should be performed for the theoretical side.

With the assumption of quark-hadron duality, we obtain
%
%%%%%%%%%%%%%%%%%%%%%%%%%%%%%%%%%%%%%%%%%%%%%%%%%%%%%%%%%%%%%%%%%%%%%%%%%%%%%%
\begin{equation}
W(n, s_0)\Big |_{Hadron} = W(n, s_0)\Big |_{OPE} \, .
\end{equation}
%%%%%%%%%%%%%%%%%%%%%%%%%%%%%%%%%%%%%%%%%%%%%%%%%%%%%%%%%%%%%%%%%%%%%%%%%%%%%%
%
The mass of the ground state can be obtained as
%
%%%%%%%%%%%%%%%%%%%%%%%%%%%%%%%%%%%%%%%%%%%%%%%%%%%%%%%%%%%%%%%%%%%%%%%%%%%%%%
\begin{equation}
M^2_Y(n, s_0)= { W(n+1, s_0) \over W(n, s_0) } \, . \label{eq:FESR}
\end{equation}
%%%%%%%%%%%%%%%%%%%%%%%%%%%%%%%%%%%%%%%%%%%%%%%%%%%%%%%%%%%%%%%%%%%%%%%%%%%%%%
%

The spectral functions $\rho^M_i(s)$ can be drawn from the Borel
transformed correlation functions shown in section~\ref{sec:svz}.
The d = 12 terms which are proportional to $1 / (q^2)^2$ do not
contribute to the function $W(n,s_0)$ of Eq.~(\ref{eq:moment}) for
$n=0$, or they have a very small contribution for $n=1$, when the
theoretical side is computed by the integral over the circle of
radius $s_0$ on the complex $q^2$ plain.

\begin{figure}[h!t]
\begin{center}
\scalebox{0.8}{\includegraphics{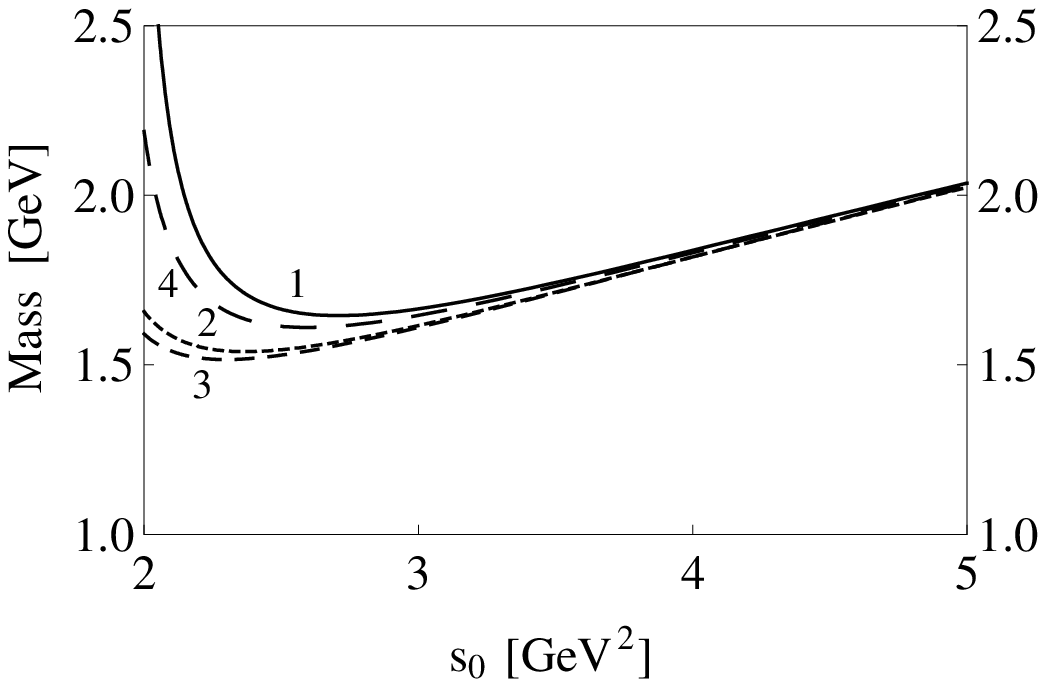}}
\scalebox{0.8}{\includegraphics{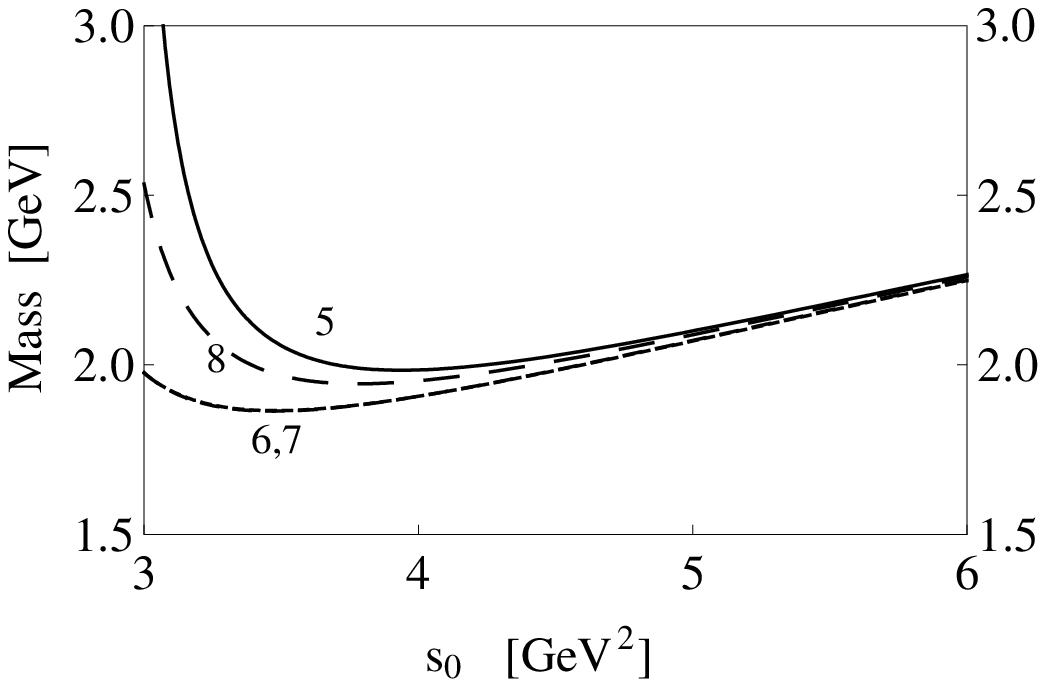}} \caption{The mass
calculated using the finite energy sum rule. The mass for the
currents $\eta^M_{1\mu}$, $\eta^M_{2\mu}$, $\eta^M_{3\mu}$ and
$\eta^M_{4\mu}$ is shown in the left hand side, and The mass for the
currents $\eta^M_{5\mu}$, $\eta^M_{6\mu}$, $\eta^M_{7\mu}$ and
$\eta^M_{8\mu}$ are shown in the right hand side. The labels besides
the lines indicate the suffix i of the current $\eta^M_{i\mu}$
($i=1,\cdots,8$).} \label{fig:fesr}
\end{center}
\end{figure}

The mass is shown as a function of the threshold value $s_0$ in
Fig.~\ref{fig:fesr}, where $n$ is chosen to be 1. We find that there
is a mass minimum. It is around 1.6 GeV for currents $\eta^M_1$,
$\eta^M_2$, $\eta^M_3$ and $\eta^M_4$, whose quark contents are $q q
\bar q \bar q$, while it is around 2.0 GeV for currents $\eta^M_5$,
$\eta^M_6$, $\eta^M_7$ and $\eta^M_8$, whose quark contents are $q s
\bar q \bar s$. Here we again find that the threshold values become
smaller than the mass obtained in the mass minimum region. See
Ref~\cite{Chen:2008ej} for details. In a short summary, we arrive at
the same results as the previous SVZ QCD sum rule.

%
%%%%%%%%%%%%%%%%%%%%%%%%%%%%%%%%%%%%%%%%%%%%%%%%%%%%%%%%%%%%%%%%%%%%%%%%%%%%%%
%---------figure 6*6
%%%%%%%%%%%%%%%%%%%%%%%%%%%%%%%%%%%%%%%%%%%%%%%%%%%%%%%%%%%%%%%%%%%%%%%%%%%%%%
%

%
%=====================================================================================
%=====================================================================================
\section{Decay Patterns of the $1^{-+}$ Tetraquark States}\label{sec:decay}
%=====================================================================================
%=====================================================================================
%

In this paper, we have verified that $(q q)(\bar q \bar q)$
construction and $(\bar q q)(\bar q q)$ construction are
equivalent (see Appendix~\ref{app:mesoncurrent}), and from the
second one we can obtain some decay information. The four
independent $(\bar q q)(\bar q q)$ currents $\xi^{M}_{i\mu}$ lead
to the same mass, and therefore, we shall study the decay patterns
from all these currents. We can obtain the $S$-wave decay patterns
straightforwardly:
\begin{enumerate}

\item The current $\xi^{M}_{1\mu}$ naively falls apart to one
scalar meson and one vector meson:
\begin{eqnarray} \label{eq:decay1}
\xi^{M}_{1\mu} &:& \pi_1(1600) \rightarrow 0^+ \left (\sigma(600),
f_0(980) \cdots \right) + 1^- \left(\rho(770),\omega(782) \cdots
\right) \, , \\ \nonumber && \pi_1(2000) \rightarrow 0^+ \left
(\sigma(600), \kappa(800) \cdots \right) + 1^- \left(\rho(770),
K^*(892) \cdots \right) \, .
\end{eqnarray}

\item The current $\xi^{M}_{2\mu}$ naively falls apart to one
axial-vector meson and one pseudoscalar meson:
\begin{eqnarray} \label{eq:decay2}
\xi^{M}_{2\mu} &:& \pi_1(1600) \rightarrow 1^+ \left(
a_1(1260),b_1(1235) \cdots \right) + 0^- \left( \pi(135) \cdots
\right) \, ,
\\ \nonumber && \pi_1(2000) \rightarrow 1^+ \left(a_1(1260), K_1(1270), \cdots \right) + 0^- \left(\pi(135), K(498) \cdots \right)  \, .
\end{eqnarray}

\item The current $\xi^{M}_{3\mu}$ naively falls apart to one
vector meson and one axial-vector meson:
\begin{eqnarray} \label{eq:decay3}
\xi^{M}_{3\mu} &:& \pi_1(1600) \rightarrow 1^- \left(
\rho(770),\omega(782) \cdots \right) + 1^+ \left(
a_1(1260),b_1(1235) \cdots \right) \, ,
\\ \nonumber && \pi_1(2000) \rightarrow 1^- \left(\rho(770), K^*(892) \cdots \right) + 1^+ \left(a_1(1260), K_1(1270) \cdots \right)  \, .
\end{eqnarray}

\item The current $\xi^{M}_{4\mu}$ naively falls apart to one
axial-vector meson and one vector meson:
\begin{eqnarray} \label{eq:decay4}
\xi^{M}_{4\mu} &:& \pi_1(1600) \rightarrow 1^+ \left(
a_1(1260),b_1(1235) \cdots \right) + 1^- \left(
\rho(770),\omega(782) \cdots \right) \, ,
\\ \nonumber && \pi_1(2000) \rightarrow 1^+ \left(a_1(1260),K_1(1270) \cdots \right) + 1^- \left(\rho(770),K^*(892) \cdots \right) \, .
\end{eqnarray}

\end{enumerate}
$\pi_1(2000)$ contains one $\bar s s$ pair, so its final states
should also contain one $\bar s s$ pair, and its decay patterns are
more complicated than $\pi_1(1600)$. We see that the decay modes
(\ref{eq:decay3}) and (\ref{eq:decay4}) are kinematically forbidden
(or strongly suppressed) due to energy conservation. The decay modes
(\ref{eq:decay1}) are difficult to be observed in the experiments
due to the large decay width of scalar mesons ($\sigma$ and
$\kappa$). Moreover, the scalar mesons below 1 GeV are sometimes
interpreted as tetraquark states, and if so, these decay modes
should be suppressed due to the extra $\bar q q$
pair~\cite{Chen:2007xr}. Therefore, the decay modes
(\ref{eq:decay2}) are preferred. The $\pi_1$ meson first decays to
one axial-vector meson and one pseudoscalar meson. Then the
axial-vector meson decays into two or more pseudoscalar mesons.
However, the second step is a $P$-wave decay. Considering the
conservation of $G$ parity, the decay mode $a_1(1260) \pi$ is
forbidden. One possible decay pattern is that $\pi_1(1600)$ first
decays to $b_1(1235) \pi$, and then decays to $\omega \pi \pi$.

We can also check the $P$-wave decay patterns besides $S$-wave
decay patterns. We find that the current $\xi^{M}_{2\mu}$ leads to
a decay mode of two $P$-wave pseudoscalar mesons by naively
relating $\bar q \gamma_\mu \gamma_5 q$ and $\partial_\mu \pi$
\begin{eqnarray} \label{eq:decayPwave}
\pi_1(1600) &\rightarrow& 0^- \left( \pi, \eta, \eta^\prime \cdots
\right) + 0^- \left( \pi, \eta, \eta^\prime \cdots \right) \, ,
\\ \nonumber \pi_1(2000) &\rightarrow& 0^- \left( \pi, \eta, \eta^\prime \cdots \right)
+ 0^- \left( \pi, \eta, \eta^\prime \cdots \right)  \, .
\end{eqnarray}
Considering the conservation of $G$ parity, decay modes $\pi \pi$
and $\eta \eta$ etc. are forbidden, and possible decay modes are
$\pi \eta$ and $\pi \eta^\prime$ etc. Summarizing the decay
patterns, there are two possible decay modes: $P$-wave many body
decay, such as $\omega \pi \pi$, and $P$-wave two body decay, such
as $\pi \eta$ and $\pi \eta^\prime$. This is partly consistent with
the experiments which observe $\pi_1(1600)$ and $\pi_1(2015)$ in the
decay modes $\pi \eta^\prime$, $\omega \pi \pi$ and $\eta \pi \pi
\pi$. However, the experiment has not observe them in the final
state $\pi \eta$. Certainly it is desired to study these decay
patterns to obtain more information on the structure of the $\pi_1$s
mesons.

%
%=====================================================================================
%=====================================================================================
\section{Summary}\label{sec:summary}
%=====================================================================================
%=====================================================================================
%

In this paper we have performed the QCD sum rule analysis of the
exotic tetraquark states with $I^GJ^{PC} = 1^-1^{-+}$. The
tetraquark currents have rich internal structure. There are
several independent currents for a given set of quantum numbers.
We have classified the complete set of independent currents and
constructed the currents in the form of either $(qq)(\bar q \bar
q)$ or $(\bar qq)(\bar qq)$. As expected, they are shown to be
equivalent by having the complete set of independent currents.
Physically, this seems to make it difficult to draw interpretation
of the internal structure such as diquark ($qq$) dominated or
meson ($\bar q \bar q$) dominated ones. Using the complete set of
the currents, one can perform an optimal analysis of the QCD sum
rule.

Somewhat complicated feature arises from the flavor structure. We
have tested all possibilities for the isovector $I =1$ states. In
the $SU(3)$ limit, there are three cases of, in the diquark
$(qq)(\bar q \bar q)$ construction, $\mathbf{6} \otimes \mathbf{\bar
6}$, $\mathbf{\bar 3} \otimes \mathbf{3}$ and $(\mathbf{\bar 3}
\otimes \mathbf{\bar 6}) \oplus (\mathbf{6} \otimes \mathbf{3})$. We
find that the former two cases can not result in meaningful sum rule
since the spectral functions become negative. On the other hand, the
mixed case $(\mathbf{\bar 3} \otimes \mathbf{\bar 6}) \oplus
(\mathbf{6} \otimes \mathbf{3})$ allows positive OPE with which we
can perform the QCD sum rule analysis. Actual currents have been
constructed in the limit of the ideal mixing where the currents are
classified by the number of the strange quarks. Hence the quark
contents are either $qq \bar q \bar q$ or $q s \bar q \bar s$.

We have then performed the SVZ and finite energy sum rules. The
resulting masses are around 1.6 GeV for $qq \bar q \bar q$, and
around 2.0 GeV for $q s \bar q \bar s$. The four independent
currents lead to the same mass and couple to a single state as shown
above. Hence one of our main conclusions is that the higher energy
states $\pi_1(1600)$ and $\pi_1(2015)$ are well compatible with the
tetraquark picture in the present QCD sum rule analysis. On the
other hand, any combination of the independent currents does not
seem to couple sufficiently to the lower mass state $\pi_1(1400)$,
which was, however, described as a hybrid state by K.~C.~Yang in
Ref.~\cite{Yang:2007cc}. He obtained a low mass around 1.26 GeV by
using the renormalization-improved QCD sum rules. The $\pi_1(1400)$
state seems somewhat special, as the experiments show the similarity
between $\pi_1(1600)$ and $\pi_1(2015)$ as well as the difference
between $\pi_1(1400)$ and the above two states, which we have
discussed in the introduction.

We have also studied their decay patterns and found that these
states can be searched for in the decay mode of the axial-vector and
pseudoscalar meson pair such as $b_1(1235)\pi$, which is sometimes
considered as the characteristic decay mode of the hybrid mesons.
The P-wave modes $\pi\eta, \pi\eta'$ are also quite important.

It is also interesting to study the partners of $\pi_1$s.
Especially, we can study the one with quark contents $u d \bar s
\bar s$, which is at the top of the flavor representation
$\mathbf{\bar {10}}$ (see Fig.~\ref{fig:tetra}). It has a mass
around 2.0 GeV, and the decay modes  are $K^+ (\bar s u) K^0 (\bar s
d)$ ($P$-wave) and $KKK$ ($P$-wave) etc. BESIII will start taking
data very soon. The search/identification of exotic mesons is one of
its important physical goals. Hopefully the dedicated experimental
programs on the exotic mesons at BESIII and JLAB in the coming years
will shed light on their existence, and then their internal
structure. More work on theoretical side is also needed. We will go
on to study other tetraquark candidates.

%
%=====================================================================================
%=====================================================================================
%=====================================================================================
\section*{Acknowledgments}
%=====================================================================================
%=====================================================================================
%=====================================================================================
%

H.X.C. is grateful for Monkasho support for his stay at the Research
Center for Nuclear Physics where this work is done. This project was
supported by the National Natural Science Foundation of China under
Grants 10625521, 10721063, the Ministry of Education of China, and
the Grant for Scientific Research ((C) No.19540297) from the
Ministry of Education, Culture, Science and Technology, Japan.

%
%=====================================================================================
%=====================================================================================
%=====================================================================================
\appendix
%=====================================================================================
%=====================================================================================
%=====================================================================================
%

%
%=====================================================================================
%=====================================================================================
\section{$(\bar q q)(\bar q q)$ Currents}\label{app:mesoncurrent}
%=====================================================================================
%=====================================================================================
%

In this appendix, we attempt to construct the tetraquark currents
using quark-antiquark ($\bar q q$) pairs. For each state containing
diquark and antidiquark having the symmetric flavor $\mathbf{6_f}
\otimes \mathbf{6_f}$, there are four $(\bar q q)(\bar q q)$
currents:
%
%%%%%%%%%%%%%%%%%%%%%%%%%%%%%%%%%%%%%%%%%%%%%%%%%%%%%%%%%%%%%%%%%%%%%%%%%%%%%%
\begin{eqnarray}\nonumber \label{def:meson_S}
%-------------------------------------xi S 1------------------------------------
\xi^S_{1\mu} &=& (\bar{q}_{3a} \gamma_\mu\gamma_5
q_{1a})(\bar{q}_{4b} \gamma_5 q_{2b}) + (\bar{q}_{3a} \gamma_5
q_{1a})(\bar{q}_{4b} \gamma_\mu\gamma_5 q_{2b}) + (\bar{q}_{3a}
\gamma_\mu\gamma_5 q_{2a})(\bar{q}_{4b} \gamma_5 q_{1b}) +
(\bar{q}_{3a} \gamma_5 q_{2a})(\bar{q}_{4b} \gamma_\mu\gamma_5
q_{1b}) \, ,
%-------------------------------------xi S 2------------------------------------
\\ \nonumber \xi^S_{2\mu} &=& (\bar{q}_{3a} \gamma^\nu q_{1a})(\bar{q}_{4b}
\sigma_{\mu\nu} q_{2b}) + (\bar{q}_{3a} \sigma_{\mu\nu}
q_{1a})(\bar{q}_{4b} \gamma^\nu q_{2b})  + (\bar{q}_{3a} \gamma^\nu
q_{2a})(\bar{q}_{4b} \sigma_{\mu\nu} q_{1b}) + (\bar{q}_{3a}
\sigma_{\mu\nu} q_{2a})(\bar{q}_{4b} \gamma^\nu q_{1b}) \, ,
%-------------------------------------xi S 3------------------------------------
\\ \nonumber \xi^S_{3\mu} &=& {\lambda_{ab}}{\lambda_{cd}}\{(\bar{q}_{3a} \gamma_\mu\gamma_5 q_{1b})(\bar{q}_{4c} \gamma_5 q_{2d}) +
(\bar{q}_{3a} \gamma_5 q_{1b})(\bar{q}_{4c} \gamma_\mu\gamma_5
q_{2d}) + (\bar{q}_{3a} \gamma_\mu\gamma_5 q_{2b})(\bar{q}_{4c}
\gamma_5 q_{1d}) + (\bar{q}_{3a} \gamma_5 q_{2b})(\bar{q}_{4c}
\gamma_\mu\gamma_5 q_{1d})\} \, ,
%-------------------------------------xi S 4------------------------------------
\\ \nonumber \xi^S_{4\mu} &=& {\lambda_{ab}}{\lambda_{cd}}\{
(\bar{q}_{3a} \gamma^\nu q_{1b})(\bar{q}_{4c} \sigma_{\mu\nu}
q_{2d}) + (\bar{q}_{3a} \sigma_{\mu\nu} q_{1b})(\bar{q}_{4c}
\gamma^\nu q_{2d}) + (\bar{q}_{3a} \gamma^\nu q_{2b})(\bar{q}_{4c}
\sigma_{\mu\nu} q_{1d}) + (\bar{q}_{3a} \sigma_{\mu\nu}
q_{2b})(\bar{q}_{4c} \gamma^\nu q_{1d})\} \, .
\end{eqnarray}
%%%%%%%%%%%%%%%%%%%%%%%%%%%%%%%%%%%%%%%%%%%%%%%%%%%%%%%%%%%%%%%%%%%%%%%%%%%%%%
%
Among these currents, only two are independent. We can verify the
following relations
%
%%%%%%%%%%%%%%%%%%%%%%%%%%%%%%%%%%%%%%%%%%%%%%%%%%%%%%%%%%%%%%%%%%%%%%%%%%%%%%
\begin{eqnarray}\nonumber
\xi^S_{3\mu} &=& - \frac{5}{3} \xi^S_{1\mu} - i \xi^S_{2\mu}\, ,
\\ \nonumber \xi^S_{4\mu} &=& 3i \xi^S_{1\mu} + \frac{1}{3}
\xi^S_{2\mu}\, .
\end{eqnarray}
%%%%%%%%%%%%%%%%%%%%%%%%%%%%%%%%%%%%%%%%%%%%%%%%%%%%%%%%%%%%%%%%%%%%%%%%%%%%%%
%
Moreover, they are equivalent to the $(qq)(\bar q \bar q)$ currents
%
%%%%%%%%%%%%%%%%%%%%%%%%%%%%%%%%%%%%%%%%%%%%%%%%%%%%%%%%%%%%%%%%%%%%%%%%%%%%%%
\begin{eqnarray}\nonumber
\psi^S_{1\mu} &=& - \frac{1}{2} \xi^S_{1\mu} + \frac{i}{2}
\xi^S_{2\mu}\, ,
\\ \nonumber \psi^S_{2\mu} &=& - \frac{3i}{2} \xi^S_{1\mu} + \frac{1}{2}
\xi^S_{2\mu}\, .
\end{eqnarray}
%%%%%%%%%%%%%%%%%%%%%%%%%%%%%%%%%%%%%%%%%%%%%%%%%%%%%%%%%%%%%%%%%%%%%%%%%%%%%%
%

For each state containing diquark and antidiquark having the
antisymmetric flavor $\mathbf{\bar 3_f} \otimes \mathbf{3_f}$, there
are also four $(\bar q q)(\bar q q)$ currents which are non-zero:
%
%%%%%%%%%%%%%%%%%%%%%%%%%%%%%%%%%%%%%%%%%%%%%%%%%%%%%%%%%%%%%%%%%%%%%%%%%%%%%%
\begin{eqnarray}\nonumber \label{def:mesonA}
%-------------------------------------xi A 1------------------------------------
\xi^A_{1\mu} &=& (\bar{q}_{3a} \gamma_\mu\gamma_5
q_{1a})(\bar{q}_{4b} \gamma_5 q_{2b}) + (\bar{q}_{3a} \gamma_5
q_{1a})(\bar{q}_{4b} \gamma_\mu\gamma_5 q_{2b}) - (\bar{q}_{3a}
\gamma_\mu\gamma_5 q_{2a})(\bar{q}_{4b} \gamma_5 q_{1b}) -
(\bar{q}_{3a} \gamma_5 q_{2a})(\bar{q}_{4b} \gamma_\mu\gamma_5
q_{1b}) \, ,
%-------------------------------------xi A 2------------------------------------
\\ \nonumber \xi^A_{2\mu} &=& (\bar{q}_{3a} \gamma^\nu q_{1a})(\bar{q}_{4b}
\sigma_{\mu\nu} q_{2b}) + (\bar{q}_{3a} \sigma_{\mu\nu}
q_{1a})(\bar{q}_{4b} \gamma^\nu q_{2b}) - (\bar{q}_{3a} \gamma^\nu
q_{2a})(\bar{q}_{4b} \sigma_{\mu\nu} q_{1b}) - (\bar{q}_{3a}
\sigma_{\mu\nu} q_{2a})(\bar{q}_{4b} \gamma^\nu q_{1b}) \, ,
%-------------------------------------xi A 3------------------------------------
\\ \nonumber \xi^A_{3\mu} &=& {\lambda_{ab}}{\lambda_{cd}}\{(\bar{q}_{3a}
\gamma_\mu\gamma_5 q_{1b})(\bar{q}_{4c} \gamma_5 q_{2d}) +
(\bar{q}_{3a} \gamma_5 q_{1b})(\bar{q}_{4c} \gamma_\mu\gamma_5
q_{2d}) - (\bar{q}_{3a} \gamma_\mu\gamma_5 q_{2b})(\bar{q}_{4c}
\gamma_5 q_{1d}) - (\bar{q}_{3a} \gamma_5 q_{2b})(\bar{q}_{4c}
\gamma_\mu\gamma_5 q_{1d})\} \, ,
%-------------------------------------xi A 4------------------------------------
\\ \nonumber \xi^A_{4\mu} &=& {\lambda_{ab}}{\lambda_{cd}}\{
(\bar{q}_{3a} \gamma^\nu q_{1b})(\bar{q}_{4c} \sigma_{\mu\nu}
q_{2d}) + (\bar{q}_{3a} \sigma_{\mu\nu} q_{1b})(\bar{q}_{4c}
\gamma^\nu q_{2d}) - (\bar{q}_{3a} \gamma^\nu q_{2b})(\bar{q}_{4c}
\sigma_{\mu\nu} q_{1d}) - (\bar{q}_{3a} \sigma_{\mu\nu}
q_{2b})(\bar{q}_{4c} \gamma^\nu q_{1d})\} \, ,
\end{eqnarray}
%%%%%%%%%%%%%%%%%%%%%%%%%%%%%%%%%%%%%%%%%%%%%%%%%%%%%%%%%%%%%%%%%%%%%%%%%%%%%%
%
where two are independent
%
%%%%%%%%%%%%%%%%%%%%%%%%%%%%%%%%%%%%%%%%%%%%%%%%%%%%%%%%%%%%%%%%%%%%%%%%%%%%%%
\begin{eqnarray}\nonumber
\xi^A_{3\mu} &=& \frac{1}{3} \xi^A_{1\mu} + i \xi^A_{2\mu}\, ,
\\ \nonumber \xi^A_{4\mu} &=& -3i \xi^A_{1\mu} - \frac{5}{3}
\xi^A_{2\mu}\, .
\end{eqnarray}
%%%%%%%%%%%%%%%%%%%%%%%%%%%%%%%%%%%%%%%%%%%%%%%%%%%%%%%%%%%%%%%%%%%%%%%%%%%%%%
%
They are equivalent to the $(qq)(\bar q \bar q)$ currents
%
%%%%%%%%%%%%%%%%%%%%%%%%%%%%%%%%%%%%%%%%%%%%%%%%%%%%%%%%%%%%%%%%%%%%%%%%%%%%%%
\begin{eqnarray}\nonumber
\psi^A_{1\mu} &=& - \frac{1}{2} \xi^A_{1\mu} + \frac{i}{2}
\xi^A_{2\mu}\, ,
\\ \nonumber \psi^A_{2\mu} &=& - \frac{3i}{2} \xi^A_{1\mu} + \frac{1}{2}
\xi^A_{2\mu}\, .
\end{eqnarray}
%%%%%%%%%%%%%%%%%%%%%%%%%%%%%%%%%%%%%%%%%%%%%%%%%%%%%%%%%%%%%%%%%%%%%%%%%%%%%%
%

For the currents which have a mixed flavor symmetry, we just show
the $(\bar q q)(\bar q q)$ currents which belong to the flavor
representation $\mathbf{\bar 3_f}\otimes\mathbf{\bar6_f}$.
%
%%%%%%%%%%%%%%%%%%%%%%%%%%%%%%%%%%%%%%%%%%%%%%%%%%%%%%%%%%%%%%%%%%%%%%%%%%%%%%
\begin{eqnarray}\nonumber \label{def:mesonM}
%-------------------------------------xi M 1------------------------------------
\xi^{ML}_{1\mu} &=& (\bar{q}_{3a} q_{1a})(\bar{q}_{4b} \gamma_\mu
q_{2b}) - (\bar{q}_{3a} \gamma_\mu q_{1a})(\bar{q}_{4b} q_{2b}) -
(\bar{q}_{3a} q_{2a})(\bar{q}_{4b} \gamma_\mu q_{1b}) +
(\bar{q}_{3a} \gamma_\mu q_{2a})(\bar{q}_{4b} q_{1b}) \, ,
%-------------------------------------xi M 2------------------------------------
\\ \nonumber \xi^{ML}_{2\mu} &=& (\bar{q}_{3a} \gamma^\mu \gamma_5 q_{1a})(\bar{q}_{4b}
\gamma_5 q_{2b}) - (\bar{q}_{3a} \gamma_5 q_{1a})(\bar{q}_{4b}
\gamma^\mu \gamma_5 q_{2b}) - (\bar{q}_{3a} \gamma^\mu \gamma_5
q_{2a})(\bar{q}_{4b} \gamma_5 q_{1b}) + (\bar{q}_{3a} \gamma_5
q_{2a})(\bar{q}_{4b} \gamma^\mu \gamma_5 q_{1b}) \, ,
%-------------------------------------xi M 3------------------------------------
\\ \nonumber \xi^{ML}_{3\mu} &=& (\bar{q}_{3a} \gamma^\nu q_{1a})(\bar{q}_{4b}
\sigma_{\mu\nu} q_{2b}) - (\bar{q}_{3a} \sigma_{\mu\nu}
q_{1a})(\bar{q}_{4b} \gamma^\nu q_{2b}) - (\bar{q}_{3a} \gamma^\nu
q_{2a})(\bar{q}_{4b} \sigma_{\mu\nu} q_{1b}) + (\bar{q}_{3a}
\sigma_{\mu\nu} q_{2a})(\bar{q}_{4b} \gamma^\nu q_{1b}) \, ,
%-------------------------------------xi M 4------------------------------------
\\ \nonumber \xi^{ML}_{4\mu} &=& (\bar{q}_{3a} \gamma^\nu \gamma_5
q_{1a})(\bar{q}_{4b} \sigma_{\mu\nu} \gamma_5 q_{2b}) -
(\bar{q}_{3a} \sigma_{\mu\nu} \gamma_5 q_{1a})(\bar{q}_{4b}
\gamma^\nu \gamma_5 q_{2b}) - (\bar{q}_{3a} \gamma^\nu \gamma_5
q_{2a})(\bar{q}_{4b} \sigma_{\mu\nu} \gamma_5 q_{1b}) +
(\bar{q}_{3a} \sigma_{\mu\nu} \gamma_5 q_{2a})(\bar{q}_{4b}
\gamma^\nu \gamma_5 q_{1b}) \, .
\end{eqnarray}
%%%%%%%%%%%%%%%%%%%%%%%%%%%%%%%%%%%%%%%%%%%%%%%%%%%%%%%%%%%%%%%%%%%%%%%%%%%%%%
%
There are also four currents which have a color
$\mathbf{8_c}\otimes\mathbf{8_c}$ structure, and they can be written
as a combination of these color $\mathbf{1_c}\otimes\mathbf{1_c}$
currents. The relations between $\phi^{ML}_{i\mu}$ and
$\xi^{ML}_{i\mu}$ are:
%
%%%%%%%%%%%%%%%%%%%%%%%%%%%%%%%%%%%%%%%%%%%%%%%%%%%%%%%%%%%%%%%%%%%%%%%%%%%%%%
\begin{eqnarray}\nonumber
\psi^{ML}_{1\mu} &=& - \frac{1}{4} \xi^{ML}_{1\mu} + \frac{1}{4}
\xi^{ML}_{2\mu} + \frac{i}{4} \xi^{ML}_{3\mu} - \frac{i}{4}
\xi^{ML}_{4\mu} \, ,
\\ \nonumber \psi^{ML}_{2\mu} &=& \frac{3i}{4} \xi^{ML}_{1\mu} + \frac{3i}{4}
\xi^{ML}_{2\mu} + \frac{1}{4} \xi^{ML}_{3\mu} + \frac{1}{4}
\xi^{ML}_{4\mu} \, ,
\\ \nonumber \psi^{ML}_{3\mu} &=& \frac{1}{4} \xi^{ML}_{1\mu} + \frac{1}{4}
\xi^{ML}_{2\mu} + \frac{i}{4} \xi^{ML}_{3\mu} + \frac{i}{4}
\xi^{ML}_{4\mu} \, ,
\\ \nonumber \psi^{ML}_{4\mu} &=& - \frac{3i}{4} \xi^{ML}_{1\mu} + \frac{3i}{4}
\xi^{ML}_{2\mu} + \frac{1}{4} \xi^{ML}_{3\mu} - \frac{1}{4}
\xi^{ML}_{4\mu} \, .
\end{eqnarray}
%%%%%%%%%%%%%%%%%%%%%%%%%%%%%%%%%%%%%%%%%%%%%%%%%%%%%%%%%%%%%%%%%%%%%%%%%%%%%%
%
We can obtain similar results for $\xi^{MR}_{i\mu}$, which belong to
the flavor representation $\mathbf{6_f}\otimes\mathbf{3_f}$ can be
obtained similarly, and the currents with $J^{PC}=1^{-+}$ are
\begin{eqnarray}
\xi^{M}_{i\mu} = \xi^{ML}_{i\mu} + \xi^{MR}_{i\mu} \, .
\end{eqnarray}

%
%=====================================================================================
%=====================================================================================
\section{Two-point Correlation Functions}\label{app:ope}
%=====================================================================================
%=====================================================================================
%

In this appendix we show the results for the Borel transformed
correlation functions as defined in Eq.~(\ref{eq:borel}). Results
for the currents $\eta^A_1$, $\eta^M_2$, $\eta^M_3$, $\eta^M_4$,
$\eta^M_6$, $\eta^M_7$ and $\eta^M_8$ are indicated by the same
upper and lower indices.

%
%%%%%%%%%%%%%%%%%%%%%%%%%%%%%%%%%%%%%%%%%%%%%%%%%%%%%%%%%%%%%%%%%%%%%%%%%%%%%%
\begin{eqnarray}\nonumber
%------------------------------\rho eta_1----------------------------------
\Pi^A_1(M_B^2) &=& \int^{s_0}_{s_<} \Bigg [ {1 \over 36848 \pi^6}
s^4 - { 17 m_s^2 \over 15360 \pi^6 } s^3 + \Big ( { \langle g_s^2 G
G \rangle \over 18432 \pi^6 } + {m_s \langle \bar q q \rangle \over
192 \pi^4} + {m_s \langle \bar s s \rangle \over 96 \pi^4} \Big )
s^2 + \Big ( - { \langle \bar q q \rangle^2 \over 72 \pi^2 } - {
\langle \bar s s \rangle^2 \over 72 \pi^2 }
\\ \nonumber && - {
\langle \bar q q \rangle \langle \bar s s \rangle \over 18 \pi^2 } +
{ m_s \langle g_s \bar q \sigma G q \rangle \over 96 \pi^4 } + { m_s
\langle g_s \bar s \sigma G s \rangle \over 192 \pi^4 } - { m_s^2
\langle g_s^2 G G \rangle \over 4608 \pi^6 } \Big ) s - { \langle
\bar q q \rangle \langle g_s \bar q \sigma G q \rangle \over 48
\pi^2 } - { \langle \bar s s \rangle \langle g_s \bar s \sigma G s
\rangle \over 48 \pi^2 }
\\ \nonumber && - { \langle \bar q q \rangle \langle g_s \bar s \sigma G s
\rangle \over 24 \pi^2 } - { \langle \bar s s \rangle \langle g_s
\bar q \sigma G q \rangle \over 24 \pi^2 } + { m_s \langle g_s^2 G G
\rangle \langle \bar q q \rangle \over 256 \pi^4} - { m_s^2 \langle
\bar q q \rangle^2 \over 12 \pi^2 } + { m_s^2 \langle \bar s s
\rangle^2 \over 48 \pi^2 } + { m_s^2 \langle \bar q q \rangle
\langle \bar s s \rangle \over 4 \pi^2 } \Bigg ] e^{-s/M_B^2} ds
\\
\nonumber && + \Big ( - { \langle g_s \bar q \sigma G q \rangle^2
\over 192 \pi^2 } - { \langle g_s \bar s \sigma G s \rangle^2 \over
192 \pi^2 } - { \langle g_s \bar q \sigma G q \rangle \langle g_s
\bar s \sigma G s \rangle \over 48 \pi^2 } - { 5 \langle g_s^2 GG
\rangle \langle \bar q q \rangle \langle \bar s s \rangle \over 864
\pi^2 } + { m_s \langle \bar q q \rangle^2 \langle \bar s s \rangle
\over 3 }
\\ \nonumber && - { 2 m_s \langle \bar q q \rangle \langle \bar s s \rangle^2
\over 9 } + { 5 m_s \langle g_s^2 GG \rangle \langle g_s \bar q
\sigma G q \rangle \over 4608 \pi^4 } + { m_s^2 \langle \bar q q
\rangle \langle g_s \bar s \sigma G s \rangle \over 12 \pi^2 } + {
m_s^2 \langle \bar s s \rangle \langle g_s \bar q \sigma G q \rangle
\over 8 \pi^2 } \Big )
\\ \nonumber && + {1 \over M_B^2}\Big( - {16
g_s^2 \langle \bar q q \rangle ^2 \langle \bar s s \rangle^2 \over
81 } + { \langle g_s^2 GG \rangle \langle \bar q q \rangle \langle
g_s \bar s \sigma G s \rangle \over 1152 \pi^2 } + { \langle g_s^2
GG \rangle \langle \bar s s \rangle \langle g_s \bar q \sigma G q
\rangle \over 1152 \pi^2 } - { m_s \langle \bar q q \rangle^2
\langle g_s \bar s \sigma G s \rangle \over 9}
\\ \nonumber && - {
m_s \langle \bar s s \rangle^2 \langle g_s \bar q \sigma G q \rangle
\over 18} - { 5 m_s \langle \bar q q \rangle \langle \bar s s
\rangle \langle g_s \bar q \sigma G q \rangle \over 18} - { m_s
\langle \bar q q \rangle \langle \bar s s \rangle \langle g_s \bar s
\sigma G s \rangle \over 18}
\\ \nonumber && + { m_s^2 \langle g_s
\bar q \sigma G q \rangle^2 \over 48 \pi^2 } + { m_s^2 \langle g_s
\bar q \sigma G q \rangle \langle g_s \bar s \sigma G s \rangle
\over 48 \pi^2 } \Big)\, .
\end{eqnarray}
%%%%%%%%%%%%%%%%%%%%%%%%%%%%%%%%%%%%%%%%%%%%%%%%%%%%%%%%%%%%%%%%%%%%%%%%%%%%%%
%

%
%%%%%%%%%%%%%%%%%%%%%%%%%%%%%%%%%%%%%%%%%%%%%%%%%%%%%%%%%%%%%%%%%%%%%%%%%%%%%%
\begin{eqnarray}\nonumber
%------------------------------\rho 1600 2----------------------------------
\Pi^M_2(M_B^2) &=& \int^{s_0}_{0} \Bigg [ {1 \over 6144 \pi^6} s^4 +
{ 11 \langle g_s^2 G G \rangle \over 18432 \pi^6 } s^2 + { \langle
\bar q q \rangle^2 \over 6 \pi^2 } s + { \langle \bar q q \rangle
\langle g_s \bar q \sigma G q \rangle \over 4 \pi^2 }  \Bigg ]
e^{-s/M_B^2} ds
\\ \nonumber && + \Big ( { \langle g_s \bar q \sigma G q \rangle^2 \over 16 \pi^2
} + { 5 \langle g_s^2 GG \rangle \langle \bar q q \rangle^2 \over
864 \pi^2 } \Big ) + {1 \over M_B^2}\Big( - {32 g_s^2 \langle \bar q
q \rangle ^4 \over 27 } - { \langle g_s^2 GG \rangle \langle \bar q
q \rangle \langle g_s \bar q \sigma G q \rangle \over 576 \pi^2 }
\Big)\, .
\end{eqnarray}
%%%%%%%%%%%%%%%%%%%%%%%%%%%%%%%%%%%%%%%%%%%%%%%%%%%%%%%%%%%%%%%%%%%%%%%%%%%%%%
%
%
%%%%%%%%%%%%%%%%%%%%%%%%%%%%%%%%%%%%%%%%%%%%%%%%%%%%%%%%%%%%%%%%%%%%%%%%%%%%%%
\begin{eqnarray}\nonumber
%------------------------------\rho 1600 3----------------------------------
\Pi^M_3(M_B^2) &=& \int^{s_0}_{0} \Bigg [ {1 \over 36864 \pi^6} s^4
+ { \langle g_s^2 G G \rangle \over 18432 \pi^6 } s^2 + { \langle
\bar q q \rangle^2 \over 36 \pi^2 } s + { \langle \bar q q \rangle
\langle g_s \bar q \sigma G q \rangle \over 24 \pi^2 } \Bigg ]
e^{-s/M_B^2} ds
\\ \nonumber && + \Big ( { \langle g_s \bar q \sigma G q \rangle^2 \over 96 \pi^2
} + { 5 \langle g_s^2 GG \rangle \langle \bar q q \rangle ^2 \over
864 \pi^2 } \Big ) + {1 \over M_B^2}\Big( - {16 g_s^2 \langle \bar q
q \rangle ^4 \over 81 } - { \langle g_s^2 GG \rangle \langle \bar q
q \rangle \langle g_s \bar q \sigma G q \rangle \over 576 \pi^2 }
\Big)\, .
\end{eqnarray}
%%%%%%%%%%%%%%%%%%%%%%%%%%%%%%%%%%%%%%%%%%%%%%%%%%%%%%%%%%%%%%%%%%%%%%%%%%%%%%
%
%
%%%%%%%%%%%%%%%%%%%%%%%%%%%%%%%%%%%%%%%%%%%%%%%%%%%%%%%%%%%%%%%%%%%%%%%%%%%%%%
\begin{eqnarray}\nonumber
%------------------------------\rho 1600 4----------------------------------
\Pi^M_4(M_B^2) &=& \int^{s_0}_{0} \Bigg [ {1 \over 12288 \pi^6} s^4
+ { \langle g_s^2 G G \rangle \over 18432 \pi^6 } s^2 + { \langle
\bar q q \rangle^2 \over 12 \pi^2 } s + { \langle \bar q q \rangle
\langle g_s \bar q \sigma G q \rangle \over 8 \pi^2 } \Bigg ]
e^{-s/M_B^2} ds
\\ \nonumber && + \Big ( { \langle g_s \bar q \sigma G q \rangle^2 \over 32 \pi^2
} - { 5 \langle g_s^2 GG \rangle \langle \bar q q \rangle^2 \over
864 \pi^2 } \Big ) + {1 \over M_B^2}\Big( - {16 g_s^2 \langle \bar q
q \rangle ^4 \over 27 } + { \langle g_s^2 GG \rangle \langle \bar q
q \rangle \langle g_s \bar q \sigma G q \rangle \over 576 \pi^2 }
\Big)\, .
\end{eqnarray}
%%%%%%%%%%%%%%%%%%%%%%%%%%%%%%%%%%%%%%%%%%%%%%%%%%%%%%%%%%%%%%%%%%%%%%%%%%%%%%
%

%
%%%%%%%%%%%%%%%%%%%%%%%%%%%%%%%%%%%%%%%%%%%%%%%%%%%%%%%%%%%%%%%%%%%%%%%%%%%%%%
\begin{eqnarray}\nonumber
%------------------------------\rho 2000 2----------------------------------
\Pi^M_6(M_B^2) &=& \int^{s_0}_{4 m_s^2} \Bigg [ {1 \over 6144 \pi^6}
s^4 - { 17 m_s^2 \over 2560 \pi^6 } s^3 + \Big (  { 11 \langle g_s^2
G G \rangle \over 18432 \pi^6 } - {m_s \langle \bar q q \rangle
\over 32 \pi^4} + {m_s \langle \bar s s \rangle \over 16 \pi^4} \Big
) s^2 + \Big ( - { \langle \bar q q \rangle^2 \over 12 \pi^2 } + {
\langle \bar q q \rangle \langle \bar s s \rangle \over 3 \pi^2 }
\\ \nonumber && - {
\langle \bar s s \rangle^2 \over 12 \pi^2 } - { m_s \langle g_s \bar
q \sigma G q \rangle \over 16 \pi^4 } + { m_s \langle g_s \bar s
\sigma G s \rangle \over 32 \pi^4 } - { 109 m_s^2 \langle g_s^2 G G
\rangle \over 18432 \pi^6 } \Big ) s - { \langle \bar q q \rangle
\langle g_s \bar q \sigma G q \rangle \over 8 \pi^2 } + { \langle
\bar q q \rangle \langle g_s \bar s \sigma G s \rangle \over 4 \pi^2
}
\\ \nonumber && + { \langle \bar s s \rangle \langle g_s \bar q
\sigma G q \rangle \over 4 \pi^2 } - { \langle \bar s s \rangle
\langle g_s \bar s \sigma G s \rangle \over 8 \pi^2 } - { 3 m_s
\langle g_s^2 G G \rangle \langle \bar q q \rangle \over 128 \pi^4}
+ { 5 m_s \langle g_s^2 G G \rangle \langle \bar s s \rangle \over
256 \pi^4} - { m_s^2 \langle \bar q q \rangle^2 \over 2 \pi^2 } - {
3 m_s^2 \langle \bar q q \rangle \langle \bar s s \rangle \over 2
\pi^2 }
\\ \nonumber && +
{ m_s^2 \langle \bar s s \rangle^2 \over 8 \pi^2 } \Bigg ]
e^{-s/M_B^2} ds + \Big ( - { \langle g_s \bar q \sigma G q \rangle^2
\over 32 \pi^2 } + { \langle g_s \bar q \sigma G q \rangle \langle
g_s \bar s \sigma G s \rangle \over 8 \pi^2 } - { \langle g_s \bar s
\sigma G s \rangle^2 \over 32 \pi^2 } - { 25 \langle g_s^2 GG
\rangle \langle \bar q q \rangle^2 \over 1728 \pi^2 } \\ \nonumber
&& + { 5 \langle g_s^2 GG \rangle \langle \bar q q \rangle \langle
\bar s s \rangle \over 144 \pi^2 } - { 25 \langle g_s^2 GG \rangle
\langle \bar s s \rangle^2 \over 1728 \pi^2 } - { 5 m_s \langle
g_s^2 GG \rangle \langle g_s \bar q \sigma G q \rangle \over 768
\pi^4 } + { 25 m_s \langle g_s^2 GG \rangle \langle g_s \bar s
\sigma G s \rangle \over 4608 \pi^4 } \\
\nonumber &&  + 2 m_s \langle \bar q q \rangle^2 \langle \bar s s
\rangle + {4  m_s \langle \bar q q \rangle \langle \bar s s
\rangle^2 \over 3 } - { m_s^2 \langle \bar q q \rangle \langle g_s
\bar s \sigma G s \rangle \over 2 \pi^2 } - { 3 m_s^2 \langle \bar s
s \rangle \langle g_s \bar q \sigma G q \rangle \over 4 \pi^2 } \Big
) \\
\nonumber && + {1 \over M_B^2}\Big( - {32 g_s^2 \langle \bar q q
\rangle ^2 \langle \bar s s \rangle^2 \over 27 }  + { 5 \langle
g_s^2 GG \rangle \langle \bar q q \rangle \langle g_s \bar q \sigma
G q \rangle \over 1152 \pi^2 } - { \langle g_s^2 GG \rangle \langle
\bar q q \rangle \langle g_s \bar s \sigma G s \rangle \over 192
\pi^2 } - { \langle g_s^2 GG \rangle \langle \bar s s \rangle
\langle g_s \bar q \sigma G q \rangle \over 192 \pi^2 } \\ \nonumber
&& + { 5 \langle g_s^2 GG \rangle \langle \bar s s \rangle \langle
g_s \bar s \sigma G s \rangle \over 1152 \pi^2 }
 - { 2 m_s \langle \bar q q \rangle^2 \langle g_s \bar s \sigma G
s \rangle \over 3} - { 5 m_s \langle \bar q q \rangle \langle \bar s
s \rangle \langle g_s \bar q \sigma G q \rangle \over 3} + { m_s
\langle \bar q q \rangle \langle \bar s s \rangle \langle g_s \bar s
\sigma G s \rangle \over 3} \\ \nonumber && + { m_s \langle \bar s s
\rangle^2 \langle g_s \bar q \sigma G q \rangle \over 3} - { 5 m_s^2
\langle g_s^2 GG \rangle \langle \bar s s \rangle^2 \over 1152 \pi^2
} + { m_s^2 \langle g_s \bar q \sigma G q \rangle^2 \over 8 \pi^2 }
- { m_s^2 \langle g_s \bar q \sigma G q \rangle \langle g_s \bar
s\sigma G s \rangle \over 8 \pi^2 } \Big)\, .
\end{eqnarray}
%%%%%%%%%%%%%%%%%%%%%%%%%%%%%%%%%%%%%%%%%%%%%%%%%%%%%%%%%%%%%%%%%%%%%%%%%%%%%%
%
%
%%%%%%%%%%%%%%%%%%%%%%%%%%%%%%%%%%%%%%%%%%%%%%%%%%%%%%%%%%%%%%%%%%%%%%%%%%%%%%
\begin{eqnarray}\nonumber
%------------------------------\rho 2000 3----------------------------------
\Pi^M_7(M_B^2) &=& \int^{s_0}_{4 m_s^2} \Bigg [ {1 \over 36864
\pi^6} s^4 - { 17 m_s^2 \over 15360 \pi^6 } s^3 + \Big ( { \langle
g_s^2 G G \rangle \over 18432 \pi^6 } - {m_s \langle \bar q q
\rangle \over 192 \pi^4} + {m_s \langle \bar s s \rangle \over 96
\pi^4} \Big ) s^2 + \Big ( - { \langle \bar q q \rangle^2 \over 72
\pi^2 } + { \langle \bar q q \rangle \langle \bar s s \rangle \over
18 \pi^2 }
\\ \nonumber && - { \langle \bar s s \rangle^2 \over 72 \pi^2 }  -
{ m_s \langle g_s \bar s \sigma G s \rangle \over 96 \pi^4 } + { m_s
\langle g_s \bar s \sigma G s \rangle \over 192 \pi^4 } - { m_s^2
\langle g_s^2 G G \rangle \over 4608 \pi^6 } \Big ) s - { \langle
\bar q q \rangle \langle g_s \bar q \sigma G q \rangle \over 48
\pi^2 } + { \langle \bar q q \rangle \langle g_s \bar s \sigma G s
\rangle \over 24 \pi^2 }
\\ \nonumber && + { \langle \bar s s \rangle \langle g_s
\bar q \sigma G q \rangle \over 24 \pi^2 } - { \langle \bar s s
\rangle \langle g_s \bar s \sigma G s \rangle \over 48 \pi^2 } - {
m_s \langle g_s^2 G G \rangle \langle \bar q q \rangle \over 256
\pi^4} - { m_s^2 \langle \bar q q \rangle^2 \over 12 \pi^2 } - {
m_s^2 \langle \bar q q \rangle \langle \bar s s \rangle \over 4
\pi^2 } + { m_s^2 \langle \bar s s \rangle^2 \over 48 \pi^2 } \Bigg
] e^{-s/M_B^2} ds
\\ \nonumber && + \Big ( - { \langle g_s \bar q \sigma G q \rangle^2 \over 192 \pi^2
} + { \langle g_s \bar q \sigma G q \rangle \langle g_s \bar s
\sigma G s \rangle \over 48 \pi^2 } - { \langle g_s \bar s \sigma G
s \rangle^2 \over 192 \pi^2 } + { 5 \langle g_s^2 GG \rangle \langle
\bar q q \rangle \langle \bar s s \rangle \over 864 \pi^2 } + { m_s
\langle \bar q q \rangle^2 \langle \bar s s \rangle \over 3 } \\
\nonumber && + { 2 m_s \langle \bar q q \rangle \langle \bar s s
\rangle^2 \over 9 } - { 5 m_s \langle g_s^2 GG \rangle \langle g_s
\bar q \sigma G q \rangle \over 4608 \pi^4 } - { m_s^2 \langle \bar
s s \rangle \langle g_s \bar q \sigma G q \rangle \over 8 \pi^2 } -
{ m_s^2 \langle \bar q q \rangle \langle g_s \bar s \sigma G
s\rangle \over 12 \pi^2 } \Big ) + {1 \over M_B^2}\Big( - {16 g_s^2
\langle \bar q q \rangle ^2 \langle \bar s s \rangle^2 \over 81 } \\
\nonumber && - { \langle g_s^2 GG \rangle \langle \bar q q \rangle
\langle g_s \bar s \sigma G s \rangle \over 1152 \pi^2 } - { \langle
g_s^2 GG \rangle \langle \bar s s \rangle \langle g_s \bar q \sigma
G q \rangle \over 1152 \pi^2 } - { m_s \langle \bar q q \rangle^2
\langle g_s \bar s \sigma G s \rangle \over 9} - {5 m_s \langle \bar
q q \rangle \langle \bar s s \rangle \langle g_s \bar q \sigma G q
\rangle \over 18} \\
\nonumber && + {m_s \langle \bar q q \rangle \langle \bar s s
\rangle \langle g_s \bar s \sigma G s \rangle \over 18} + {m_s
\langle \bar s s \rangle^2 \langle g_s \bar q \sigma G q \rangle
\over 18} + {m_s^2 \langle g_s \bar q \sigma G q \rangle^2 \over 48
\pi^2} - {m_s^2 \langle g_s \bar q \sigma G q \rangle \langle g_s
\bar s \sigma G s \rangle \over 48 \pi^2} \Big)\, .
\end{eqnarray}
%%%%%%%%%%%%%%%%%%%%%%%%%%%%%%%%%%%%%%%%%%%%%%%%%%%%%%%%%%%%%%%%%%%%%%%%%%%%%%
%
%
%%%%%%%%%%%%%%%%%%%%%%%%%%%%%%%%%%%%%%%%%%%%%%%%%%%%%%%%%%%%%%%%%%%%%%%%%%%%%%
\begin{eqnarray}\nonumber
%------------------------------\rho 2000 4----------------------------------
\Pi^M_8(M_B^2) &=& \int^{s_0}_{4 m_s^2} \Bigg [ {1 \over 12288
\pi^6} s^4 - { 17 m_s^2 \over 5120 \pi^6 } s^3 + \Big (  { \langle
g_s^2 G G \rangle \over 18432 \pi^6 } - {m_s \langle \bar q q
\rangle \over 64 \pi^4} + {m_s \langle \bar s s \rangle \over
32\pi^4} \Big ) s^2 + \Big ( - { \langle \bar q q \rangle^2 \over 24
\pi^2 } + { \langle \bar q q \rangle \langle \bar s s \rangle \over
6 \pi^2 }
\\ \nonumber && - { \langle \bar s s \rangle^2 \over 24 \pi^2 } -
{ m_s \langle g_s \bar q \sigma G q \rangle \over 32 \pi^4 } + { m_s
\langle g_s \bar s \sigma G s \rangle \over 64 \pi^4 } - { 17 m_s^2
\langle g_s^2 G G \rangle \over 18432 \pi^6 } \Big ) s - { \langle
\bar q q \rangle \langle g_s \bar q \sigma G q \rangle \over 16
\pi^2 } + { \langle \bar q q \rangle \langle g_s \bar s \sigma G s
\rangle \over 8\pi^2 } \\ \nonumber && + { \langle \bar s s \rangle
\langle g_s \bar q \sigma G q \rangle \over 8\pi^2 } - { \langle
\bar s s \rangle \langle g_s \bar s \sigma G s \rangle \over 16
\pi^2 } + { m_s \langle g_s^2 G G \rangle \langle \bar s s \rangle
\over 256 \pi^4} - { m_s^2 \langle \bar q q \rangle^2 \over 4 \pi^2
} - { 3 m_s^2 \langle \bar q q \rangle \langle \bar s s \rangle
\over 4 \pi^2 } + { m_s^2 \langle \bar s s \rangle^2 \over 16 \pi^2
} \Bigg ] e^{-s/M_B^2} ds
\\ \nonumber && + \Big ( - { \langle g_s \bar q \sigma G q \rangle^2 \over 64 \pi^2
} + { \langle g_s \bar q \sigma G q \rangle \langle g_s \bar s
\sigma G s \rangle \over 16 \pi^2 } - { \langle g_s \bar s \sigma G
s \rangle^2 \over 64 \pi^2 } - { 5 \langle g_s^2 GG \rangle \langle
\bar q q \rangle^2 \over 1728 \pi^2 } - { 5 \langle g_s^2 GG \rangle
\langle \bar s s \rangle^2 \over 1728 \pi^2 } \\
\nonumber && + { 5 m_s \langle g_s^2 GG \rangle \langle g_s \bar s
\sigma G s \rangle \over 4608 \pi^4 } + m_s \langle \bar q q
\rangle^2 \langle \bar s s \rangle  + { 2 m_s \langle \bar q q
\rangle \langle \bar s s^2 \rangle \over 3 } - { 3 m_s^2 \langle
\bar s s \rangle \langle g_s \bar q \sigma G q \rangle \over 8 \pi^2
} - { m_s^2 \langle \bar q q \rangle \langle g_s \bar s \sigma G s
\rangle
\over 4 \pi^2 } \Big ) \\
\nonumber && + {1 \over M_B^2}\Big( - {16 g_s^2 \langle \bar q q
\rangle ^2 \langle \bar s s \rangle^2 \over 27 } + { \langle g_s^2
GG \rangle \langle \bar q q \rangle \langle g_s \bar q \sigma G q
\rangle \over 1152 \pi^2 } + { \langle g_s^2 GG \rangle \langle \bar
s s \rangle \langle g_s \bar s \sigma G s \rangle \over 1152 \pi^2 }
- { m_s \langle \bar q q \rangle^2 \langle g_s \bar s \sigma G s
\rangle \over 3} \\ \nonumber && - { 5 m_s \langle \bar q q \rangle
\langle \bar s s \rangle \langle g_s \bar q \sigma G q \rangle \over
6} + { m_s \langle \bar q q \rangle \langle \bar s s \rangle \langle
g_s \bar s \sigma G s \rangle \over 6} + { m_s \langle \bar s s
\rangle^2 \langle g_s \bar q \sigma G q \rangle \over 6} - { m_s^2
\langle g_s^2 GG \rangle \langle \bar s s \rangle^2 \over 1152 \pi^2
} \\ \nonumber && + {m_s^2 \langle g_s \bar q \sigma G q \rangle^2
\over 16 \pi^2} - {m_s^2 \langle g_s \bar q \sigma G q \rangle
\langle g_s \bar s \sigma G s \rangle \over 16 \pi^2} \Big)\, .
\end{eqnarray}
%%%%%%%%%%%%%%%%%%%%%%%%%%%%%%%%%%%%%%%%%%%%%%%%%%%%%%%%%%%%%%%%%%%%%%%%%%%%%%
%

%%%%%%%%%%%%%%%%%%%%%%%%%%%%%%%%%%%%%%%%%%%%%%%%%%%%%%%%%%%%%%%%%%%%%%%%%%%%%%

%%%%%%%%%%%%%%%%%%%%%%%%%%%%%%%%%%%%%%%%%%%%%%%%%%%%%%%%%%%%%%%%%%%%%%%%%%%%%%
%

\end{document}